\documentclass[aps,prx,letter,preprintnumbers,superscriptaddress,nofootinbib,longbibliography,floatfix]{revtex4-2}
\pdfoutput=1
\usepackage{rotating}
\usepackage{array}
\usepackage{amsmath}
\usepackage[normalem]{ulem}
\usepackage{slashed}
\usepackage{booktabs}
\usepackage[pdftex,table]{xcolor}
\usepackage{units}
\usepackage{xfrac}
\usepackage{mathtools}
\usepackage{empheq}
\usepackage[]{units}
\usepackage{multirow}
\usepackage{amssymb}
\usepackage{url}
\usepackage{comment}
\usepackage{physics}
\usepackage{color,soul}
\usepackage{bbm}
\usepackage{adjustbox}
\usepackage[T1]{fontenc}
\usepackage{xspace}

\usepackage{graphicx,caption,subcaption,tikz}
\usetikzlibrary{backgrounds}

\usetikzlibrary{calc}
\DeclareCaptionFormat{overlay}{\gdef\capoverlay{#1#2#3\par}}
\DeclareCaptionStyle{overlay}{format=overlay}
\DeclareCaptionFormat{suboverlay}{\gdef\subcapoverlay{(\thesubfigure) #3\par}}
\DeclareCaptionStyle{suboverlay}{format=suboverlay}

\usepackage{hyperref}
\hypersetup{
  colorlinks=true,
  citecolor=blue,
  linkcolor=blue,
  urlcolor=blue
}

\newcommand{\ours}{\textsc{Parnassus}\xspace}

\begin{document}

\title{Conditional Deep Generative Models for\\Simultaneous Simulation and Reconstruction of Entire Events}

\author{Etienne Dreyer}
\affiliation{Weizmann Institute of Science, Rehovot, Israel}

\author{Eilam Gross}
\affiliation{Weizmann Institute of Science, Rehovot, Israel}

\author{Dmitrii Kobylianskii}
\email{dmitry.kobylyansky@weizmann.ac.il}
\affiliation{Weizmann Institute of Science, Rehovot, Israel}

\author{Vinicius Mikuni}
\affiliation{Lawrence Berkeley National Laboratory, Berkeley, CA 94720, USA}

\author{Benjamin Nachman}
\email{bpnachman@lbl.gov}
\affiliation{Lawrence Berkeley National Laboratory, Berkeley, CA 94720, USA}

\begin{abstract}
We extend the Particle-flow Neural Assisted Simulations (\textsc{Parnassus}) framework of fast simulation and reconstruction to entire collider events.  In particular, we use two generative Artificial Intelligence (genAI) tools, continuous normalizing flows and diffusion models, to create a set of reconstructed particle-flow objects conditioned on truth-level particles from CMS Open Simulations.  While previous work focused on jets, our updated methods now can accommodate all particle-flow objects in an event along with particle-level attributes like particle type and production vertex coordinates.  This approach is fully automated, entirely written in Python, and GPU-compatible.  Using a variety of physics processes at the LHC, we show that the extended \textsc{Parnassus} is able to generalize beyond the training dataset and outperforms the standard, public tool \textsc{Delphes}.

\end{abstract}

\maketitle

\section{Introduction}

Particle physics is in a precision era where experiments can be digitally emulated from sub-nuclear scales all the way to the macroscopic size of detectors.  Simulated datasets are an integral part of interpreting complex experimental data.  Reconstruction algorithms reduce the dimensionality of the inference task by combining real or simulated measurements on many detector components into estimated particle properties.  These properties are then the basis for a multitude of downstream analysis tasks.  As detectors become more complex and datasets become ever larger, the computational challenges of first-principles simulations and state-of-the-art reconstruction algorithms will become prohibitive~\cite{Elvira:2022wyn}.

To address this challenge, a suite of fast simulation programs have been developed.  Experimental collaborations have developed tailored approaches like those used by ATLAS~\cite{ATLAS:2010arf,ATLAS:2010bfa,ATLAS:2021pzo,ATLAS:2022jhk} and CMS~\cite{Abdullin:2011zz,Giammanco:2014bza,Sekmen:2016iql} at the LHC.  While many pieces of these programs use classical sampling methods, there is a growing interest in generative artificial intelligence (GenAI) for the slowest parts of the simulations, notably calorimeters~\cite{Krause:2024avx}.  These fast surrogate models mimic the output of full detector simulations and are thus processed using the same reconstruction algorithms as the data. While this approach offers dramatic speedup for simulation, it does not alleviate the significant computational cost of reconstruction~\cite{CERN-LHCC-2022-005,Software:2815292}.

Outside of experimental collaborations, generic tools like \textsc{Delphes}~\cite{deFavereau:2013fsa,Selvaggi:2014mya,Mertens:2015kba} have combined fast simulation with fast reconstruction.  While widely-used and highly impactful for phenomenological studies, the precision of \textsc{Delphes} is fundamentally limited because it operates by smearing input particles and the smearing functions are hard-coded based on published performance results.  Recently, there have been proposals to address these and other related challenges with GenAI.  \textsc{FlashSim}~\cite{2704573,Vaselli:2024vrx} uses GenAI to generate reconstructed jets and other high-level objects conditioned on the corresponding particle-level inputs.  \textsc{Parnassus}~\cite{fastsim,Kobylianskii:2024sup,Dreyer:2024bhs} uses similar tools, operating at the level of individual (reconstructed) particles.  These approaches are automatically tunable, can accommodate non-trivial reconstruction effects, and are additionally lightweight, having only Python dependencies and being natively GPU-compatible.  \textsc{Parnassus} is proposed to be a general-purpose tool that has a similar output as \textsc{Delphes}.  Previous studies with \textsc{Parnassus} have focused on hadronic jets, which have many constituent particles with complex substructure.  However, a complete fast simulation and reconstruction tool should also be able to accommodate entire events.

\begin{figure}[h!]
    \centering
    \includegraphics[width=0.95\linewidth]{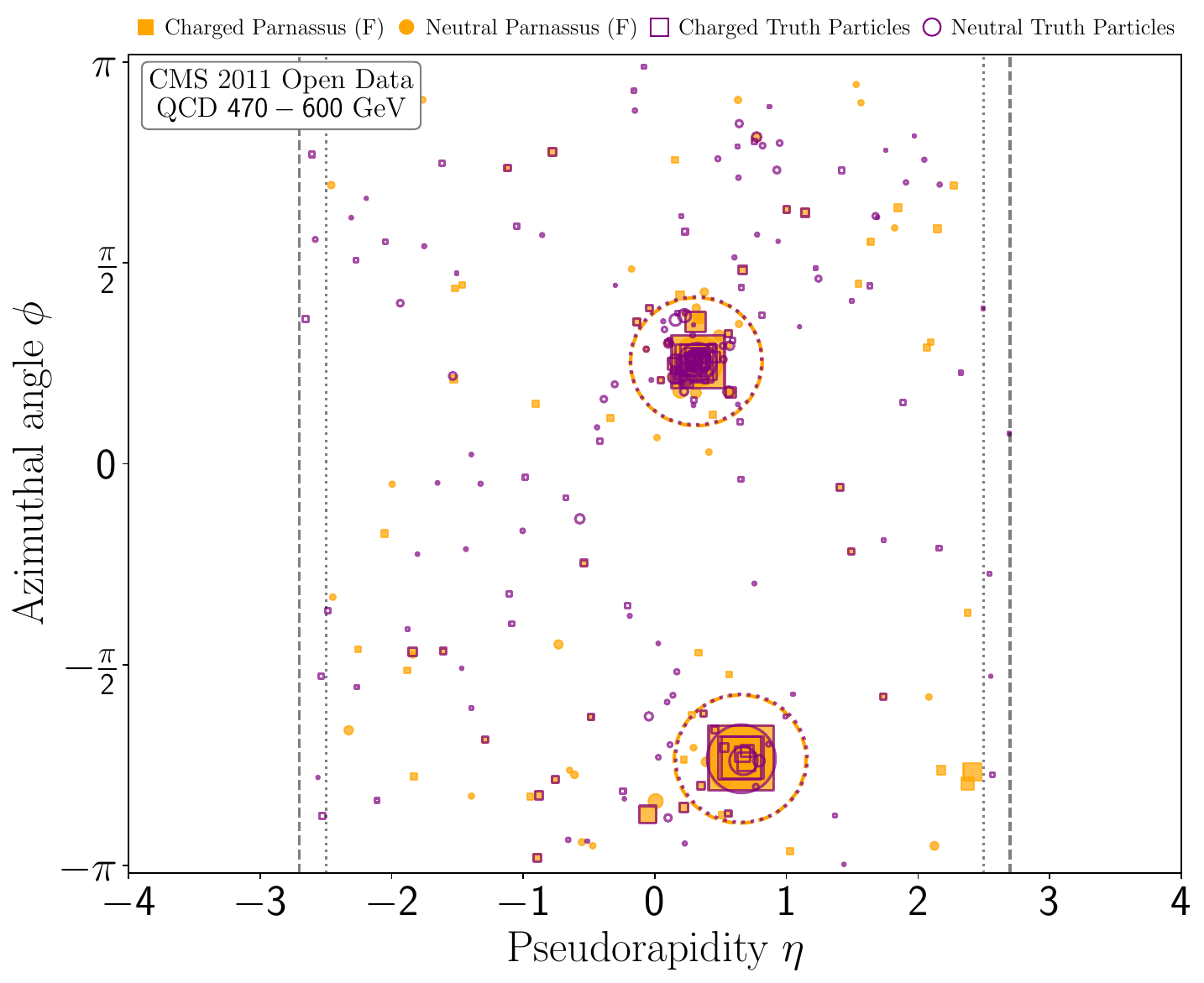}
    \caption{An event display of a single dijet event from the CMS Open Simulation. The two highest-$p_\mathrm{T}$ jets are illustrated with dashed and dotted circles while all detector-stable and reconstructed particles are shown with markers. The size of the marker reflects the particle energy. The dashed vertical lines denote the training cut $\abs{\eta} < 2.7$, and dotted lines denote the tracker acceptance area $\abs{\eta} < 2.5$.  While there is no correct answer due to the stochastic nature of the detector response, this display is a qualitative demonstration \textsc{Parnassus}'s ability to capture the full event. }
    \label{fig:eventdisplay}
\end{figure}

\begin{figure}[h!]
    \centering
    \begin{subcaptionblock}[t]{.3\textwidth}
        \includegraphics[width=\linewidth]{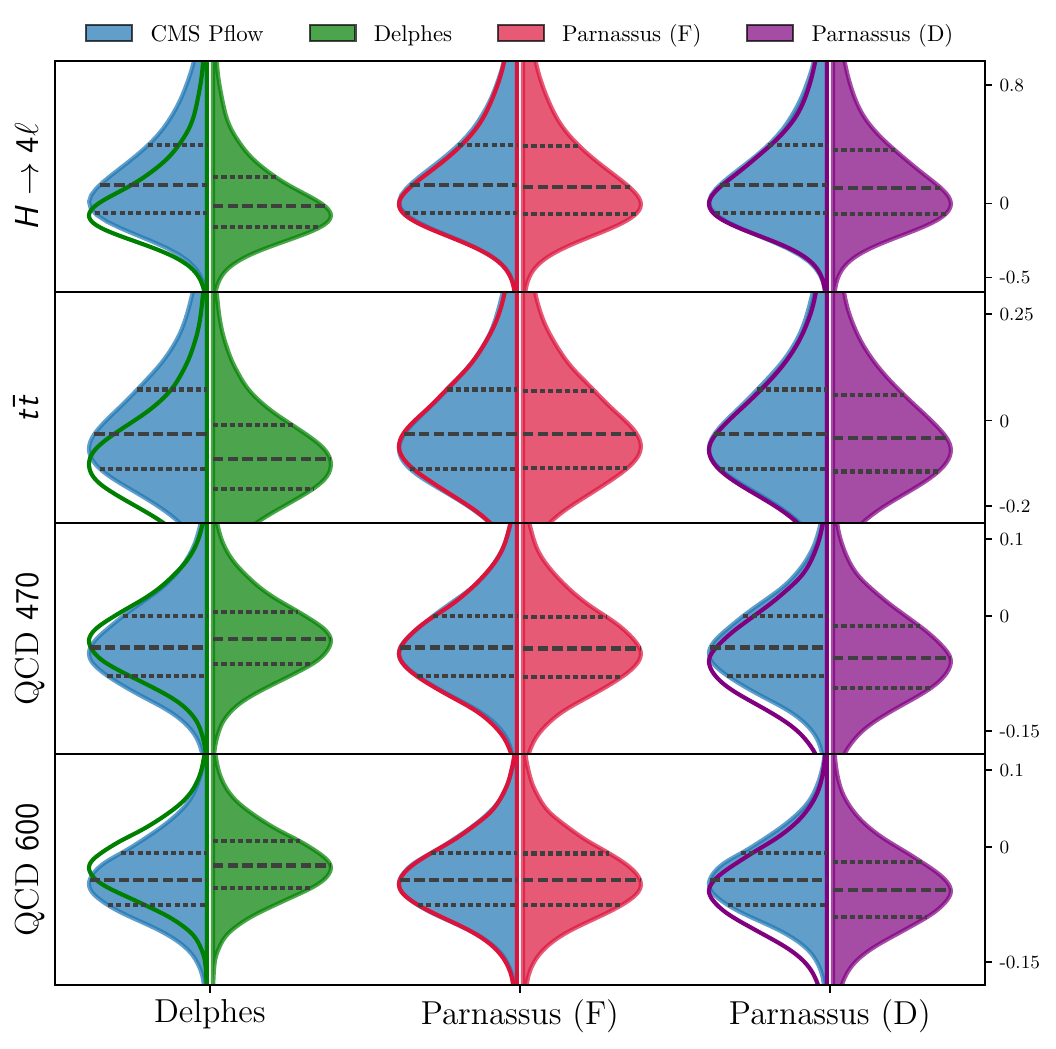}
        \caption{\footnotesize $H_T$ residuals}
    \end{subcaptionblock}
    \begin{subcaptionblock}[t]{.3\textwidth}
        \includegraphics[width=\linewidth]{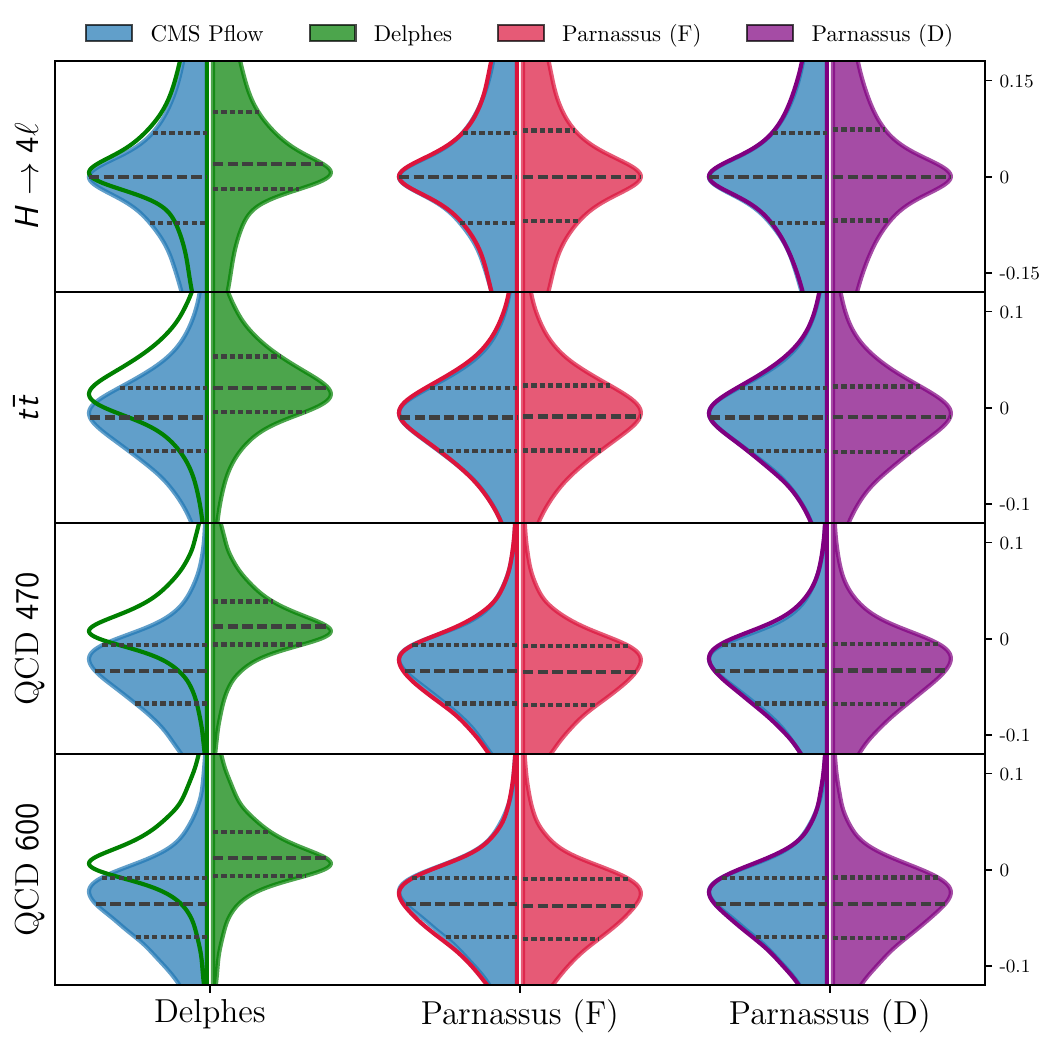}
        \caption{\footnotesize Jet $C_2$ residuals}
    \end{subcaptionblock}
    \begin{subcaptionblock}[t]{.3\textwidth}
        \includegraphics[width=\linewidth]{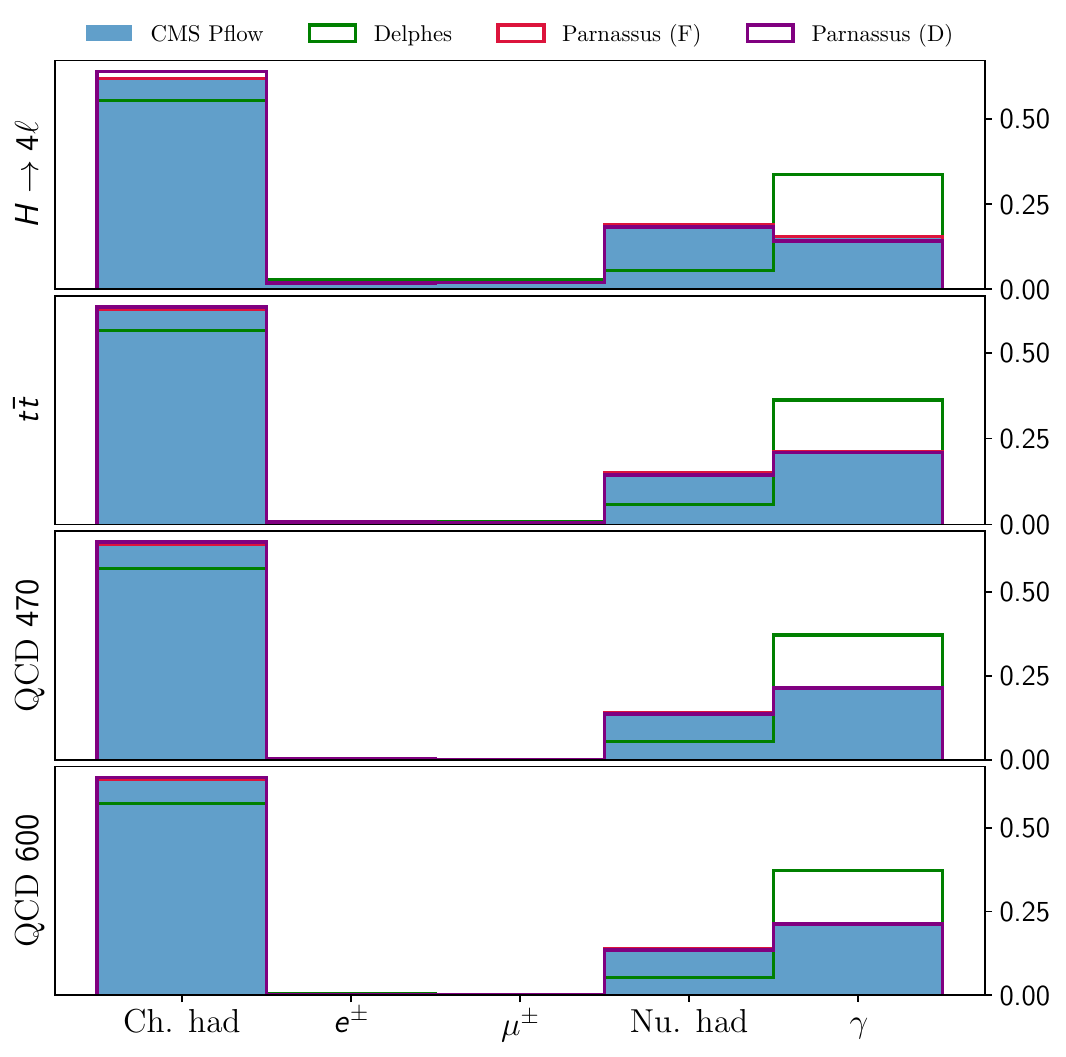}
        \caption{\footnotesize Reconstructed particle classes}
    \end{subcaptionblock}
    \caption{Three metrics summarizing the performance of the target CMS full simulation in blue compared to \textsc{Delphes} in green and \textsc{Parnassus} in red (Flow Matching) and purple (Diffusion). (a) Violin plots of the fractional residual for the scalar sum of all particle $p_\mathrm{T}$ in the event ($H_\mathrm{T}$), (b) Violin plots of the fractional residual for the leading jet $C_2$ jet substructure observable, and (c) histograms of the reconstructed particle types. Residuals are computed using the difference between the truth and the reconstructed quantities from each algorithm. In all cases, the red and purple distributions agree well with the blue, better than the green ones do.}
    \label{fig:overview}
\end{figure}

In this paper, we extend \textsc{Parnassus} to full events.  While we focus on collider physics, the methods can be applied to any experiment whose data are in an event format and are composed of (reconstructed) particles.  While much of the complication from collider events at high energy originates with jets, there are new challenges that arise when extending the model to full events.  For example, entire events have no obvious anchor to center all of the particles (like the jet axis).  Similarly, the identification of $-\pi$ and $+\pi$ in azimuthal angle is now relevant.  Furthermore, the overall multiplicity increases with a wide range of energy and length scales that need to be modeled.  We address these challenges by extending the neural network models from our previous works.  Figures~\ref{fig:eventdisplay} and~\ref{fig:overview} present high-level summaries of the new method capabilities, with many more detailed results available later in the manuscript.

This paper is organized as follows.  Section~\ref{sec:data} introduces the CMS Open Simulation dataset for our study.  The details of the machine learning methods are introduced in Sec.~\ref{sec:methods}.  Numerical results are presented in Sec.~\ref{sec:results} and the paper ends with conclusions and outlook in Sec.~\ref{sec:conclusions}.

\vspace{15mm}

\clearpage

\section{Dataset}
\label{sec:data}
We use a suite of physics processes in order to build and test the universality of our model.  Selected processes include $t\bar{t}$, $gg\rightarrow H\to 4\ell$, and QCD $2\rightarrow 2$ as these were available with high statistics in the 2011 CMS Open Simulation. Events were generated with \textsc{Pythia} 6.4.25 \cite{Sjostrand:2006za} using the CMS detector simulation based on \textsc{Geant}4 \cite{GEANT4:2002zbu} and then the events were reconstructed using the CMS particle-flow algorithm~\cite{CMS:2017yfk}. We preprocessed these datasets from the AODSIM format, keeping the set of particle-flow candidates (PFCs) and generator-level final state truth particles\footnote{Detector-stable particles with \textsc{Pythia}~6 status code 1.} (GENs). 
Each GEN or PFC particle is described by its transverse momentum, pseudorapidity, azimuthal angle, production vertex coordinates, and class ($p_\mathrm{T}, \eta, \phi, \vec{v}, \text{class}$). Particles are categorized into one of five classes: charged hadron, electron, muon, neutral hadron, or photon.

To constrain our model within the tracker acceptance area ($\abs{\eta} < 2.5$), we apply a relaxed pseudorapidity cut of $\abs{\eta} < 2.7$ for both PFC and GEN particles. During performance evaluation, we enforce a strict cut of $\abs{\eta} < 2.5$. This strategy allows \ours to learn the impact of truth particles outside the tracker on those inside.  To mitigate tracking inefficiencies and improve momentum reconstruction, we follow Refs.~\cite{komiske_exploring_2020, Tripathee_2017} and require $p_\mathrm{T} > 1$ GeV for PFCs. In order to reduce the cardinality of truth particles, we also require them to have $p_\mathrm{T} > 0.25$ GeV.
\begin{table}[h!]
    \centering
    \begin{tabular}{ccccc}
        \toprule
         Dataset & $p_\mathrm{T}^{\text{min}}$ - $p_\mathrm{T}^{\text{max}}$ [GeV]  & Type & Training & Testing\\\hline
         $H\to 4\ell$~\cite{h4lep_data} & - & Out-of-distribution &  & $\checkmark$ \\
         $t\bar{t}$~\cite{ttbar_data} & - & In distribution & $\checkmark$ & $\checkmark$ \\
         QCD~\cite{qcd470_data} & $470-600$ & In distribution & $\checkmark$ & $\checkmark$ \\
         QCD~\cite{qcd600_data} & $600-800$ & Out-of-distribution &  & $\checkmark$ \\
         \bottomrule
    \end{tabular}
    \caption{Information about used MC event samples. Usage and parton-level $p_\mathrm{T}$ ranges for QCD samples are shown.}
    \label{tab:samples}
\end{table}

We used a total of four datasets, with two dedicated to model training and in-distribution performance evaluation and the remaining two for out-of-distribution checks (see \autoref{tab:samples}). Each training dataset consisted of 1.4M events, while each testing dataset comprised 50k events, resulting in a total of 2.8M events for training and 200k events for testing.

The goal of \textsc{Parnassus} is to take as input the GEN particles and produce PFCs so that the distribution of PFCs matches that of the full CMS detector simulation and reconstruction.  As a baseline for comparison, we also create a dataset with \textsc{Delphes}~3.5.  In order to eliminate statistical fluctuations, we use exactly the same GEN events for \textsc{Delphes} by converting the GEN events from the CMS dataset into \textsc{HepMC}3 format~\cite{Buckley_2021}.  We then pass these events through \textsc{Delphes} with the default CMS detector card.  As the CMS datasets do not have truth information for the pileup, the pileup is treated as noise (e.g. the model needs to produce pileup PFCs without a starting GEN pileup particle).  To match the CMS setup, we add minimum bias events to the \textsc{Delphes} samples following the procedure in Ref.~\cite{parnassus}, with the mean number of pileup vertices per event set to 6.35.  We disable smearing of the truth primary vertex in \textsc{Delphes} in order to match the full simulation, while the pileup vertices are smeared with the default values.

\section{Methods}
\label{sec:methods}

The core of our approach is a point cloud (set of PFCs) generative model that is conditioned on another point cloud (GEN particles).  We consider two state-of-the-art methods for achieving this aim -- one based on flow matching (\textsc{Parnassus} (F)) and one based on diffusion models (\textsc{Parnassus} (D)) -- as described below.

\subsubsection{Flow and Diffusion}
Continuous normalizing flows (CNF)~\cite{chen2019neural} are a competitive generative modeling method.  Like other methods, CNFs map a base probability density $p_0$ to a target probability density $p_1$.  In the case of CNFs, this mapping is defined by a vector field $v_\theta$, parameterized by a neural network, that solves an ordinary differential equation:
\begin{equation}
    \mathrm{d}x = v_{\theta}(t, x) \mathrm{d}t,
\end{equation}
with $t\in[0, 1]$.  A common approach to learning $v_\theta$ is through a regression task called Flow Matching (FM)~\cite{lipman_flow_2023}:
\begin{equation}
\label{eq:FM}
    \mathcal{L}_{\text{FM}}(\theta) = \mathbb{E}_{t,p_t(x)}||v_{\theta}(t, x) - u_t(x)||^2\,,
\end{equation}
where $\mathcal{L}$ is the loss function for optimizing the parameters $\theta$ of the neural network and the function $u_t$ describes the target probability flow field mapping $p_0$ to $p_1$.  Since $u_t$ is not generally known in closed form, a surrogate objective is to predict an alternative form, $u_t(x|z)$, using the Conditional Flow Matching (CFM) loss function~\cite{lipman_flow_2023,albergoBuildingNormalizingFlows2023,liuFlowStraightFast2023}:
\begin{equation}
    \mathcal{L}_{\text{CFM}}(\theta) = \mathbb{E}_{t, q(z),p_t(x|z)}||v_{\theta}(t, x) - u_t(x|z)||^2,
    \label{eq:cfm_loss}
\end{equation}
where the sample-conditional approach uses $z \sim q(z), x \sim p_t(x | z)$.
Yao et al.~\cite{yaoFasterDiTFasterDiffusion2024} and Ke et al.~\cite{ke_proreflow_2025} showed that additionally steering the direction of the velocity network using a cosine similarity loss term enhances the model's convergence rate:
\begin{equation}
    \mathcal{L}_{\text{sim}}(\theta) = \mathbb{E}_{t, q(z),p_t(x|z)}\left(1 - \frac{v_{\theta}(t, x) \cdot u_t(x|z)}{\left|v_{\theta}(t, x)\right|\left|u_t(x|z)\right|}\right),
    \label{eq:cos_loss}
\end{equation}
In the previous study~\cite{parnassus}, we identified the condition $z$ with the single sample $x_\ast$ (in that case, a set of PFCs in the jet) and the target probability path was defined as:
\begin{equation}
    \begin{aligned}
        p_t(x|z) &= \mathcal{N}(x | tx_\ast, (t\sigma-t+1)^2 \mathrm{I})\,,\\
        u_t(x|z) &= \frac{x_\ast - (1 - \sigma)x}{1 - (1-\sigma)t}\,,
    \end{aligned}
    \label{eq:prob_path}
\end{equation}
which represents a path between a standard normal distribution and a Gaussian distribution centered at $x_\ast$ with standard deviation $\sigma$, which we took to be equal to $10^{-4}$.

In contrast, in this work, we identify $z$ with a pair of random variables, one from the data distribution and the other from a standard normal distribution. The path between them is defined by a family of affine Gaussian probability paths, which can be expressed as:
\begin{equation}
    \begin{aligned}
        p_t(x | x_\ast) &= \mathcal{N}(x|\alpha_t x_\ast, \sigma_t^2 \mathrm{I)}\\
        u_t(x| x_\ast) &= \dot{\alpha}_tx_t + \dot{\sigma}_t\varepsilon
        \label{eq:prob_path_affine}
    \end{aligned}
\end{equation}
where $\varepsilon \sim \mathcal{N}(0, \mathrm{I})$ and $\alpha_t, \sigma_t: [0, 1] \to [0, 1]$ are smooth functions of $t$ satisfying:\footnote{Contrary to Ref.~\cite{lipman_flow_2023} we inverted time, as in Ref.~\cite{ma_sit_2024}.}
\begin{equation}
    \alpha_0 = \sigma_1 = 1, \alpha_1 = \sigma_0 = 0, \quad -\dot{\alpha}_t, \dot{\sigma}_t > 0 \text{ for } t \in (0, 1),
\end{equation}
This case also subsumes probability paths generated by standard diffusion models~\cite{Song2021ScoreBasedGM,karras_elucidating_2022} and the denoising score matching (SM) loss can be obtained from the CFM loss (\autoref{eq:cfm_loss}) by reparameterization~\cite{lipman_flow_2024}:
\begin{equation}
    \mathcal{L}_{\text{SM}}(\theta) = \mathbb{E}_{t,\varepsilon,x_t}||s_{\theta}(t, x_t) - \nabla\log p_t(x_t)||^2,
    \label{eq:sm_loss}
\end{equation}
where $\nabla \log p_t(x_t) = -\frac{\epsilon}{\sigma(t)}$.
Salimans and Ho in Ref.~\cite{salimans2022progressive} proposed another parametrization of score matching loss, via $\mathrm{v}$-prediction, allowing more stable training:\footnote{To avoid confusion with the flow vector field $v$ we denote it by straight $\mathrm{v}$.}
\begin{equation}
    \mathcal{L}_{\text{v}}(\theta) = \mathbb{E}_{t,\varepsilon,x_t}||\mathrm{v}_{\theta}(t, x_t) - \mathrm{v}_t(x_t)||^2\,,
    \label{eq:sm_v_loss}
\end{equation}
where $\mathrm{v} = \alpha_t\varepsilon - \sigma_t x_\ast$.

Data generation with the trained neural network can be done by solving either a probability flow ODE or a reverse-time SDE:
\begin{align}
    \text{ODE: } \qquad & dx_t = v_\theta(t, x_t) dt \,, \\
    \text{SDE: } \qquad & dx_t = [f(t)x_t - g^2(t)s_\theta(t, x_t)]dt + g(t)d\bar{w}\,,
\end{align}
where 
\begin{equation}
    f(t) = \frac{\dot{\alpha_t}}{\alpha_t}, \qquad g^2(t) = -2\frac{\dot{\alpha}_t\sigma_t^2 - \alpha_t\sigma_t\dot{\sigma}_t}{\alpha_t}
\end{equation}
The exact choice of affine path, training loss, and sampling procedure is summarized in~\autoref{tab:params}.
\begin{table}[h]
    \centering
    \begin{tabular*}{0.7\linewidth}{@{\extracolsep{\fill}}lcccc}
    \toprule
    \textbf{Model type} & $\alpha_t$ & $\sigma_t$ & Loss & Reverse process \\
    \midrule
    Parnassus (F) & $1 - t$ & $t$ & $\mathcal{L}_{\text{CFM}} + \mathcal{L}_{\text{sim}}$ & ODE\\
    Parnassus (D) & $\cos\frac{\pi t}{2}$ & $\sin\frac{\pi t}{2}$ & $\mathcal{L}_{\text{v}}$ & ODE\\
    \bottomrule
    \end{tabular*}
    \caption{\ours models parametrization.}
    \label{tab:params}
\end{table}

\subsubsection{Architecture Description}
We divide the prediction of PFC properties into two stages. The first stage generates event-level quantities: missing energy, the scalar sum of transverse momenta, and the number of PFCs, represented as $\mathcal{E}^{\text{pf}} = (E_{x}^{\text{miss}}, E_{y}^{\text{miss}}, H_\mathrm{T}, N_{\text{part.}})$.  The corresponding GEN quantities are denoted $\mathcal{E}^{\text{gen}}$. The second stage predicts the PFCs and their properties, represented as $\mathcal{P}^{\text{pf}} = (p_\mathrm{T}^{\text{rel}}, \eta, \phi, \vec{v}, \text{class})$, where $p_\mathrm{T}^{\text{rel}} = \frac{p_\mathrm{T}}{H_T}$, $\vec{v}$ is the production vertex, and class is the particle type.  The corresponding GEN quantities are called $\mathcal{P}^{\text{gen}}$.

Both stages are conditioned on the set of GEN particles and the GEN-level event quantities. The second stage is additionally conditioned on the PFC-level event quantities generated in the first stage. Both the GEN and PFC particle sets are ordered by $p_\mathrm{T}^{\text{rel}}$, and their features are normalized using precomputed means and standard deviations.  We consider up to 400 PFCs/GENs which is rarely exceeded.  The class feature is one-hot encoded.

The CFM and diffusion models slightly differ in their setup of these two network components:

\vspace{4mm}

\noindent \textbf{CFM Architecture}

\begin{itemize}
    \item \textbf{Event-level network}\\
    We use a ResNet~\cite{he2016deep}-like model to parameterize $v_\theta^{\mathcal{E}}$. The network processes noisy event quantities $\mathcal{E}^{\text{pf}}$, sampled according to \autoref{eq:prob_path}, as well as a set of truth particles with their features $\mathcal{P}^{\text{gen}}$, along with event scaling information and $\mathcal{E}^{\text{gen}}$ serving as global features.

    The model architecture is illustrated in \autoref{fig:event_net}. Initially, the current timestep is embedded using sine positional encoding followed by a multilayer perceptron (MLP). Global and truth particle features, together with features of the noisy $\mathcal{E}^{\text{pf}}$, are embedded into the same hidden dimension using separate MLPs. Next, the pooled representation of the truth particles is concatenated with the updated global and timestep embeddings to form the context ($c$). The updated noisy features are then processed through an adaptive-layer normalization (adaLN-zero) procedure~\cite{peebles_scalable_2023} using $c$, followed by a sequence of residual blocks. Finally, $v_\theta^{\mathcal{E}}$ is computed via an MLP conditioned on the context $c$.
    \begin{figure}[h]
        \centering
        \includegraphics[width=0.8\textwidth]{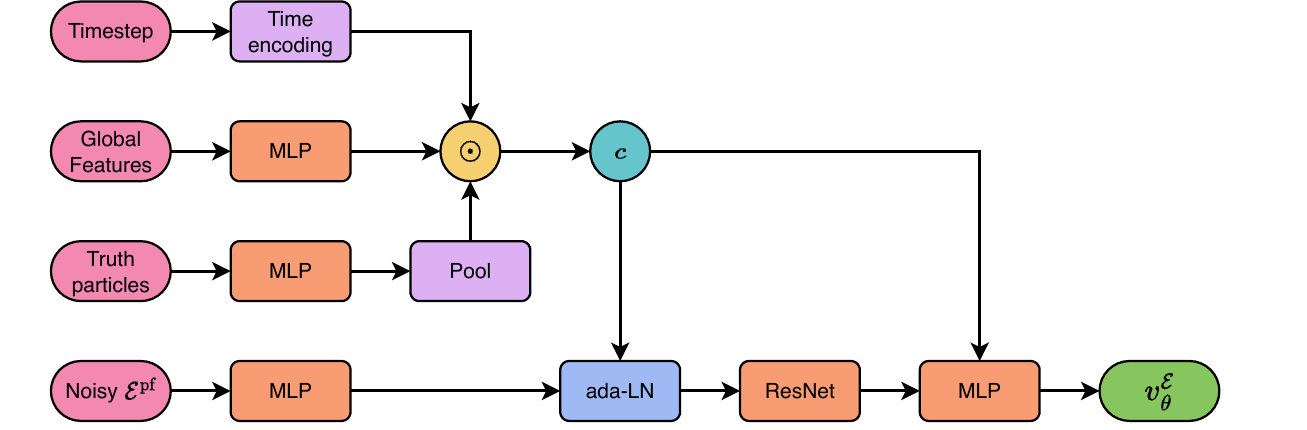}
        \caption{\textbf{Architecture of Event-level CFM model.} Concatenation is indicated by $\odot$.}
        \label{fig:event_net}
    \end{figure}

    \item \textbf{PFC-level network}
    For the PFC-level network, we employ a transformer-like model, similar to that described in Ref.~\cite{parnassus}, to parameterize $v_\theta^{\mathcal{P}}$. We also split $\phi$ into 2 variables $(\sin\phi,\cos\phi)$ to better model the neighborhood of $\pm \pi$. The network receives inputs consisting of a fixed-length set of PFCs with $\mathcal{P}^{\text{pf}}$ sampled according to \autoref{eq:prob_path_affine}, as well as a set of truth particles with $\mathcal{P}^{\text{gen}}$ along with event scaling information, $\mathcal{E}^{\text{gen}}$ and $\mathcal{E}^{\text{pf}}$ serving as global features. 
    
    \begin{figure}[h!]
        \centering
        \includegraphics[width=0.8\textwidth]{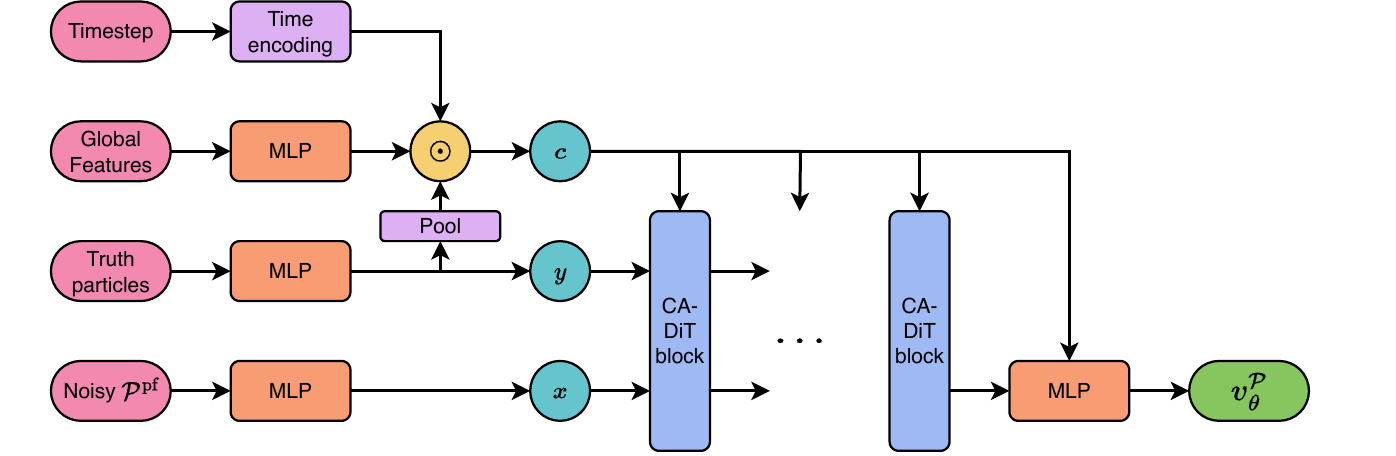}
        \caption{\textbf{Architecture of PFC-level CFM model.} Concatenation is indicated by $\odot$.}
        \label{fig:part_net}
    \end{figure}

    The model architecture, depicted in \autoref{fig:part_net}, begins with an initialization similar to that of the event stage. Following this initialization, the hidden representations of PFCs ($x$) and GEN particles ($y$) are passed through multiple CA-DiT blocks (Cross-Attention Diffusion Transformer)~\cite{parnassus, peebles_scalable_2023}, where the context $c$ is used as input to the adaptive-layer normalization (adaLN-zero)\cite{peebles_scalable_2023} in each block. 
\end{itemize}

\vspace{4mm}

\noindent \textbf{Diffusion Model Architecture}

\begin{itemize}
    \item \textbf{Event-level network}\\
    Similarly to the CFM architecture, we break down the generation into two components, resulting in two models. The event-level network uses the same architecture of the CFM model to generate the event level quantities. The main difference is that instead of generating the overall particle multiplicity as a single number, we generate instead the individual multiplicities corresponding to each particle class, with the overall multiplicity to be generated calculated from the sum over classes. The advantage of this strategy is that we do not need to individually generate the particle class using the PFC-level network, but instead we can already condition the generation of particle flow candidates based on the expected class\footnote{To be more explicit, during generation, if we need to generate two electrons, we concatenate the initial Gaussian values of two arbitrary particles with the one-hot encoded value assigned to electrons, requiring the model to generate only the remaining kinematic properties through the reverse diffusion process.}. The drawback of this approach, as we see later in the results, is that the overall multiplicity per PID becomes less accurate. Since the number of features to generate increases, we also increase the internal network representation compared to the CFM event-level network, resulting in more trainable parameters.

    \item \textbf{PFC-level network} \\
    The generation of particles is conceptually similar to the CFM model, where we create a graph- and transformer-based architecture that takes as inputs the information from generator level particles and event information to guide the diffusion process towards the generation of reconstructed particle candidates. In this case, the architecture used is inspired by \textsc{OmniLearn}~\cite{Mikuni:2024qsr,Mikuni:2025tar}. The timestep and global event information are initially embed using the same strategy as in the CFM model. The particle information at generation and reconstruction level is first passed through two DGCNN~\cite{wang2019dynamic} blocks, where $k$-nearest neighbors with $k = 3$ is used to aggregate the information of nearby particles in $\eta-\phi$ space. This new embedding is then used as inputs to four DiT~\cite{peebles_scalable_2023} blocks, where $c$ is again used as inputs to the adaLN operations. The resulting architecture is shown in Fig.~\ref{fig:pfc_net_diffusion}.
    
    \begin{figure}[h]
        \centering
        \includegraphics[width=0.9\textwidth]{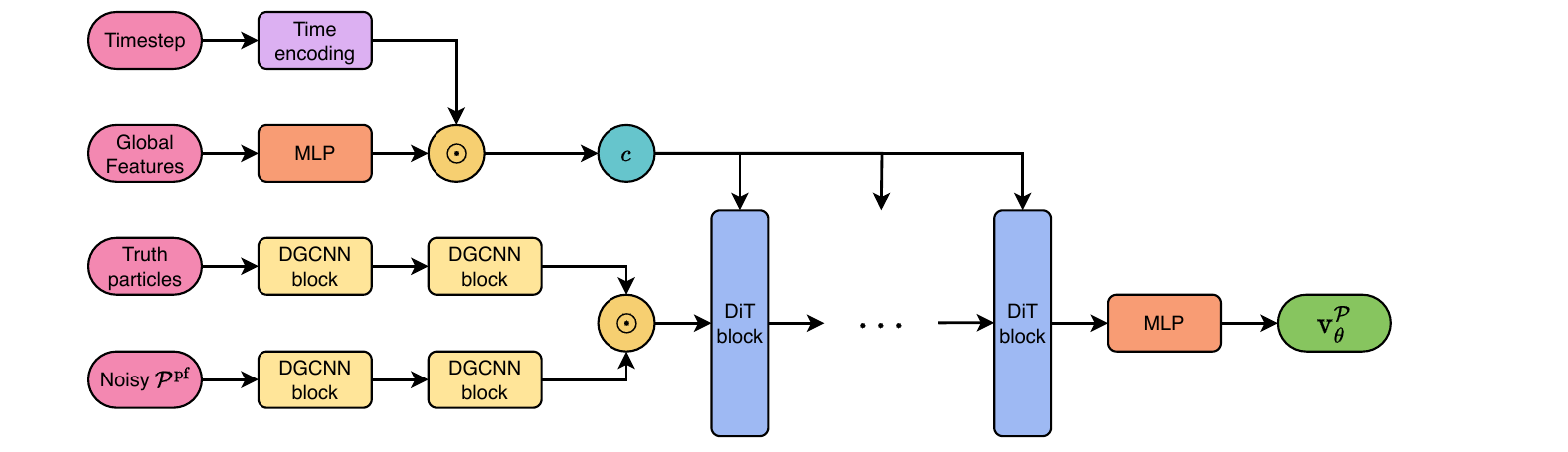}
        \caption{\textbf{Architecture of PFC-level Diffusion model.} Concatenation is indicated by $\odot$.}
        \label{fig:pfc_net_diffusion}
    \end{figure}
    
\end{itemize}
The networks for the CFM training are implemented in \textsc{PyTorch}~\cite{pytorch_my} while the diffusion model is implemented using \textsc{TensorFlow}~\cite{tensorflow}. Hyperparameters for all models are shown at~\autoref{tab:hyperparams}. The learning rate is modified according to cosine annealing schedule~\cite{loshchilov2017sgdrstochasticgradientdescent}. We also note that for the CFM model, the number of time steps can be reduced to 25 almost without losing quality.

\begin{table}[h]
    \centering
    \begin{tabular*}{0.9\linewidth}{@{\extracolsep{\fill}}lcc}
    \toprule
    \textbf{Hyperparameters} & \textbf{CFM} & \textbf{Diffusion} \\
    \midrule
    Batch size & 128 & 64 \\
    Optimizer & Lion~\cite{chenSymbolicDiscoveryOptimization2023} & Lion~\cite{chenSymbolicDiscoveryOptimization2023} \\
    Weight decay & 0.01 & 0.01 \\
    Max learning rate & $1.2 \cdot10^{-4}$ & $0.6 \cdot10^{-4}$ \\
    Min learning rate & $10^{-5}$ & $10^{-7}$ \\
    Gradient norm clip & 1.0 & 0.0 \\
    \# of epochs & 70 & 300 \\
    Sampling method & Flow-DPM-Solver~\cite{xieSANAEfficientHighResolution2024} & DPM-Solver-2~\cite{luDPMSolverFastODE2022} \\
     \# of time steps (model evaluations) & 40 (40) & 64 (128) \\
    Max output particles & 400 & 400 \\
    Trainable parameters of Event-level network & 470,532 & 1,377,352 \\
    Trainable parameters of PFC-level network & 4,864,427 & 4,101,703 \\
    \bottomrule
    \end{tabular*}
    \caption{Network hyperparameters of the \ours models.}
    \label{tab:hyperparams}
\end{table}

\clearpage

\section{Results}
\label{sec:results}

We evaluate the performance of \ours on both event-level and jet-level quantities.  Jets are clustered using FastJet~\cite{Cacciari:2011ma} with the anti-$k_t$ algorithm~\cite{Cacciari:2008gp} and $R = 0.5$.  To investigate the quality of jet properties, we match PFC jets to GEN jets using the Hungarian matching algorithm~\cite{HUNGARIAN}, with $\Delta R$ as a cost function. Since the Hungarian algorithm considers all possible matching, we additionally forbid matches with $\Delta R > 0.2$. To investigate the quality of PFC property generation, we split the PFCs into three sets: all particles, particles assigned to jets, and particles not assigned to jets (henceforth, ``background particles''). Inside each category, we additionally perform the same Hungarian-based matching between PFC and GEN particles, forbidding matches with $\Delta R > 0.6$.

We present results at three levels: event, jet, and particle. First, we examine the performance of \ours for the processes on which it was trained. For this purpose, we use a hold-out test set that is statistically identical to the training data. Second, we consider the processes that the model did not see during the training.

\begin{figure}[h]
    \centering
    \includegraphics[width=0.95\textwidth]{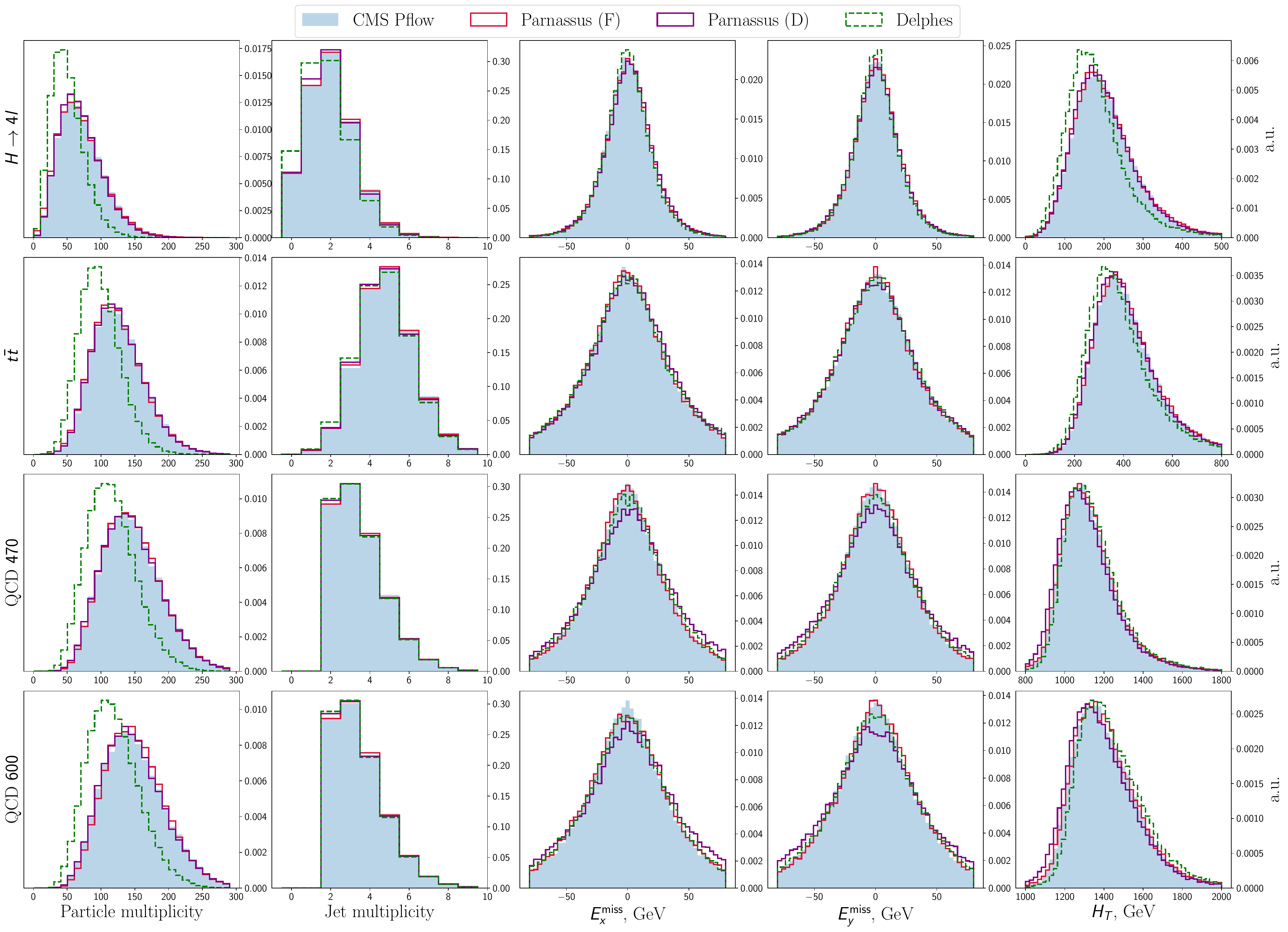}
    \caption{Histograms of event-level features for $t\bar{t}$, $H\to 4l$ and QCD datasets.  Rows correspond to different processes while columns correspond to different observables.  All histograms are normalized to unity.}
    \label{fig:evt}
\end{figure}

\subsection{Event-level performance}
As event-level metrics, we consider the number of particles, number of jets, missing transverse energy in the $x$ and $y$ directions, and the scalar sum of particle $p_\mathrm{T}$. The spectra of these observables are presented in Fig.~\ref{fig:evt}.  As all subsequent figures follow a similar pattern, we will describe this figure in detail here and then refrain from doing so for the following figures.  Each column of Fig.~\ref{fig:evt} corresponds to a different observable and each row represents a different physics process. The second and third rows show processes that were part of the training although the results shown are from independent test sets.  The first and final rows represent processes that were not part of the training.  All histograms are normalized to unity.  The filled histograms show the CMS full simulation and reconstruction - the target for both \textsc{Parnassus} and \textsc{Delphes}. The two variants of \textsc{Parnassus} are indicated with an (F) for the flow matching and (D) for diffusion.  The goal is for the unfilled histograms to be as close as possible to the filled histograms and for this agreement to span columns (observables) and rows (universality).  For the event-level observables, we can see that $H\rightarrow 4l$ has the lowest particle and jet multiplicity as well as the narrowest missing energy distribution.  In contrast, all of the other processes have multiple, hard jets (more in $t\bar{t}$ than for generic QCD dijets).  The \textsc{Parnassus} methods agree well with CMS across all processes and observables while \textsc{Delphes} is accurate for most of the distributions, but not for the particle multiplicity and $H_T$.

\begin{figure}[h]
    \centering
    \includegraphics[width=0.95\textwidth]{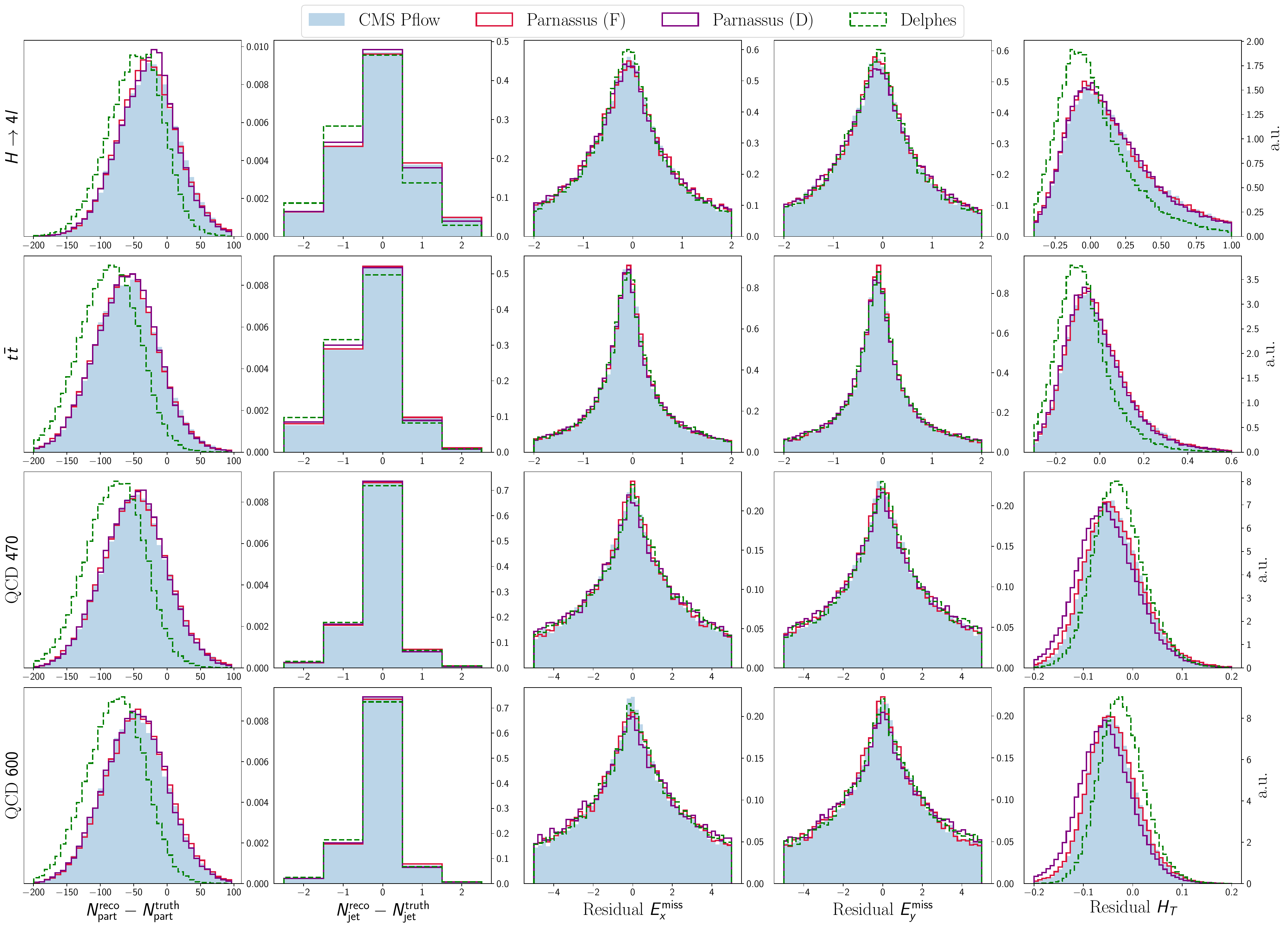}
    \caption{Histograms of the residuals of event-level features for $t\bar{t}$, $H\to 4l$ and QCD datasets. Rows correspond to different processes while columns correspond to different observables.  All histograms are normalized to unity.}
    \label{fig:evt_res}
\end{figure}

The spectra in Fig.~\ref{fig:evt} do not tell the full story, since much of these distributions are determined by the particle level distributions and not by resolution effects.  Figure~\ref{fig:evt_res} shows the residuals of the observables from Fig.~\ref{fig:evt}, to remove the dependence on the truth quantities (which are the same for all methods).  For the discrete quantities (particle and jet multiplicities), the residuals are the difference between the true and reconstructed quantities.  For the continuous quantities, the residual is the difference divided by the truth value.  The mismodeling in the particle multiplicity observed in Fig.~\ref{fig:evt} is even more pronounced in the residual.  This is also true for $H_\mathrm{T}$.  As the total particle multiplicity (and the number of jets in $H\rightarrow 4l$) is highly sensitive to detector effects beyond just smearing (e.g. fakes), which are not explicitly modeled by \textsc{Delphes}, it is not surprising that these are the observables where the relative advantage with genAI are most pronounced.

\clearpage

\subsection{Jet-level performance}

\begin{figure}[h]
    \centering
    \includegraphics[width=0.95\textwidth]{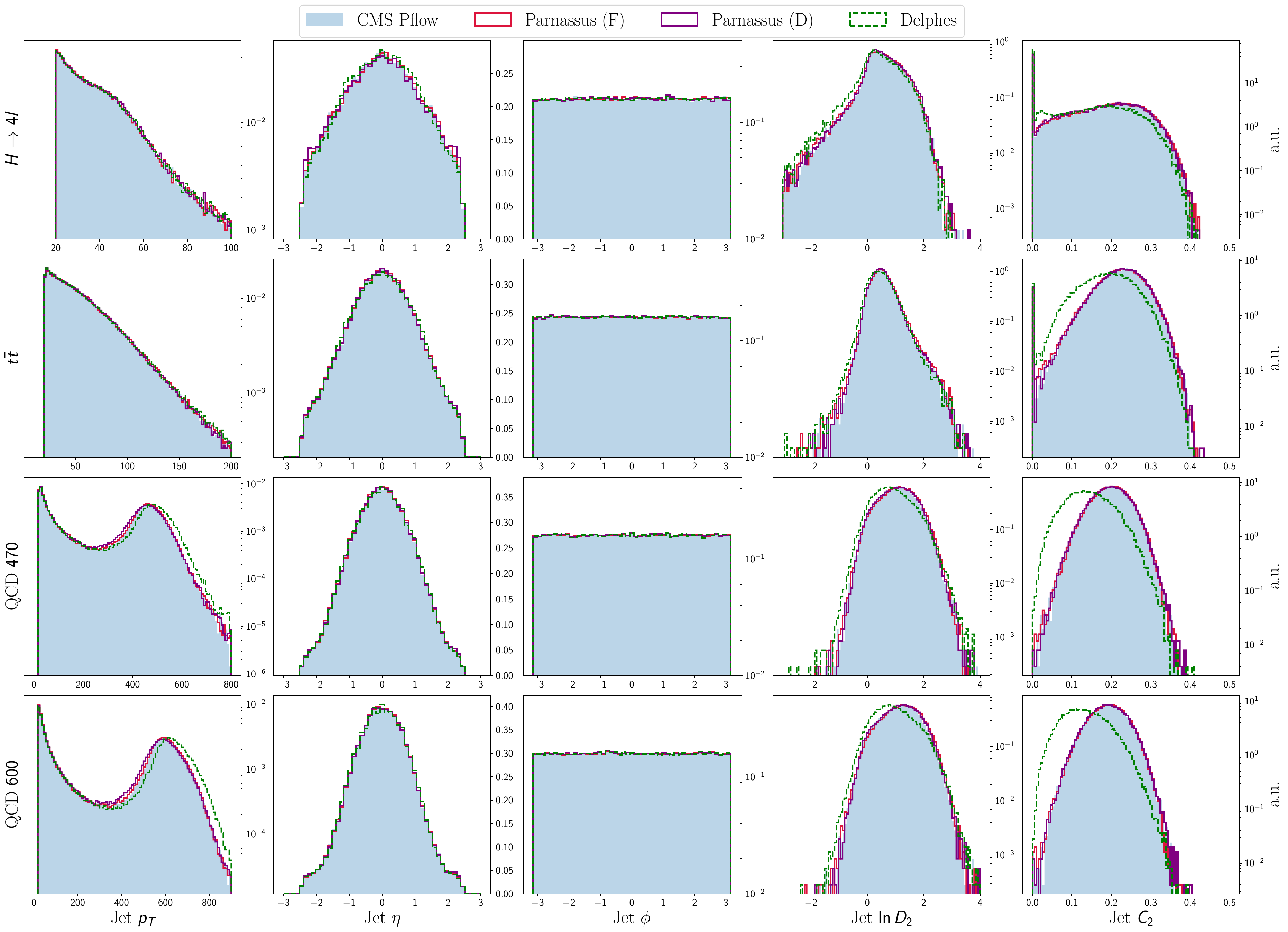}
    \caption{Histograms of jet-level features for $t\bar{t}$, $H\to 4l$ and QCD datasets. Rows correspond to different processes while columns correspond to different observables.  All histograms are normalized to unity. }
    \label{fig:jet}
\end{figure}

For the jet-level performance we explore the kinematic properties ($p_\mathrm{T}, \eta, \phi$) as well as substructure quantities, $C_2$~\cite{Larkoski:2013eya} and $D_2$~\cite{Larkoski:2014gra}.  These latter two observables characterize how consistent the radiation pattern within a jet is with a two-prong hypothesis compared to a one-prong origin.  They are widely used for boosted $W/Z/H$ boson tagging. Since we want to accurately reproduce all jets in the event, including low-$p_\mathrm{T}$ ones, we include all jets in the following figures, except for the jet substructure observables, where we restrict to the leading jets.  Figure~\ref{fig:jet} shows spectra of the jet-level quantities.  The jet $p_\mathrm{T}$ is steeply falling for the Higgs boson and top quark pair processes and bimodal for QCD.  This is because the QCD events are composed of two high $p_\mathrm{T}$ jets (the second peak) in addition to a number of additional jets superimposed (first peak).  The jet kinematic properties are mostly determined by the particle-level quantities and agree well across methods.  In contrast, the jet substructure observables are subject to significant detector effects and are much more accurate for \textsc{Parnassus} compared to \textsc{Delphes}.

\begin{figure}[h]
    \centering
    \includegraphics[width=0.95\textwidth]{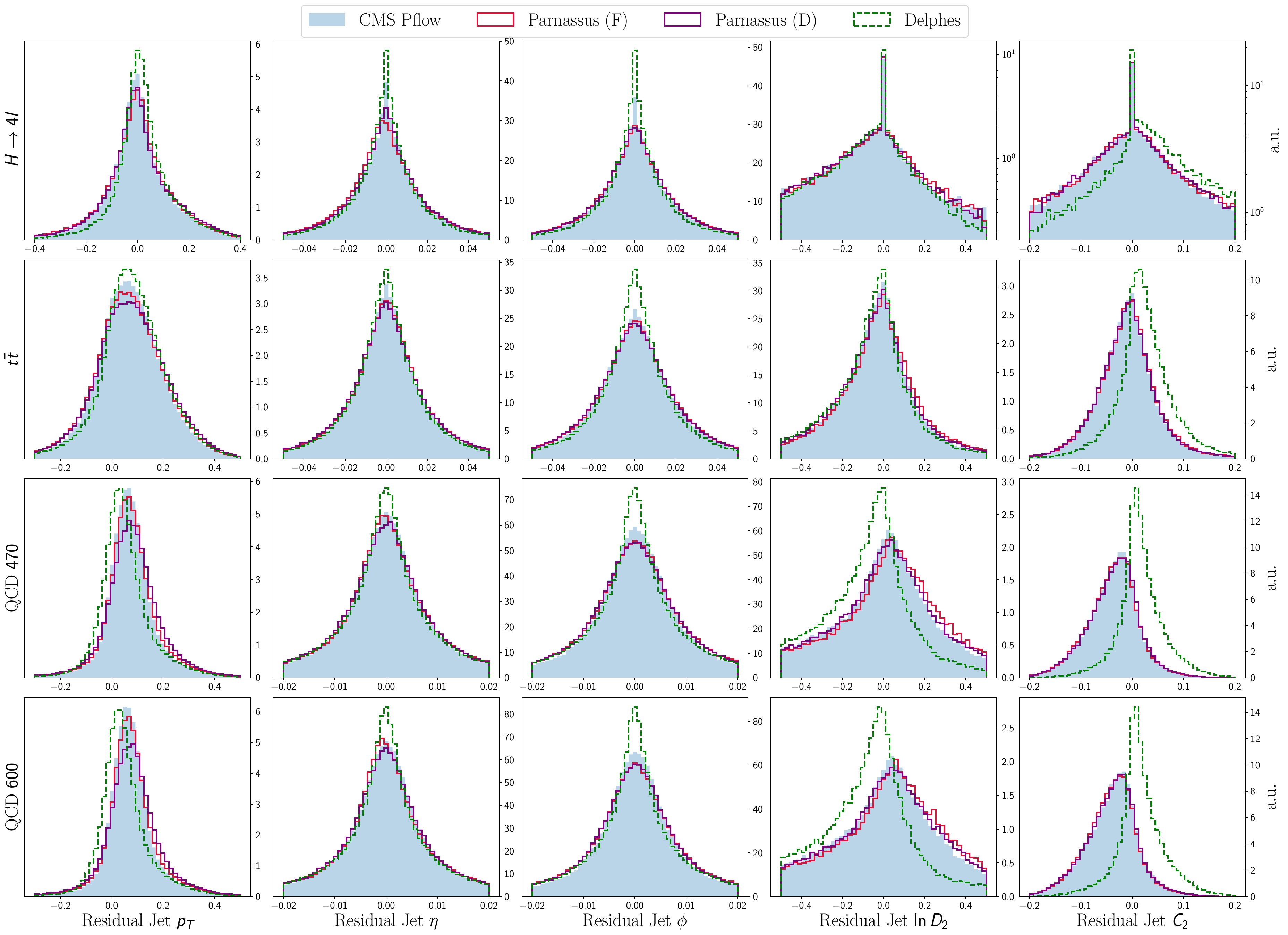}
    \caption{Histograms of residuals of jet-level features for $t\bar{t}$, $H\to 4l$ and QCD datasets. Rows correspond to different processes while columns correspond to different observables.  All histograms are normalized to unity.}
    \label{fig:jet_res}
\end{figure}

The corresponding residuals for jets are shown in \autoref{fig:jet_res}. These residuals are calculated between PFC jets and the matched GEN jet. Jets without a match are not included here.  Across the board, \textsc{Delphes} presents a narrower resolution than the CMS response, which is well-captured by both \textsc{Parnassus} models.  For the angular coordinates, the difference between \textsc{Delphes} and the other models is a pure variance effect, while there is also a bias for the other observables.  This is most prominent for $C_2$.  The \textsc{Parnassus} models also capture the improvement in resolutions of kinematic properties with $p_\mathrm{T}$, seen most clearly by comparing the third and fourth rows of \autoref{fig:jet_res}.

\clearpage

\subsection{Particle-level performance}

\begin{figure}[h!]
    \centering
    \includegraphics[width=0.95\textwidth]{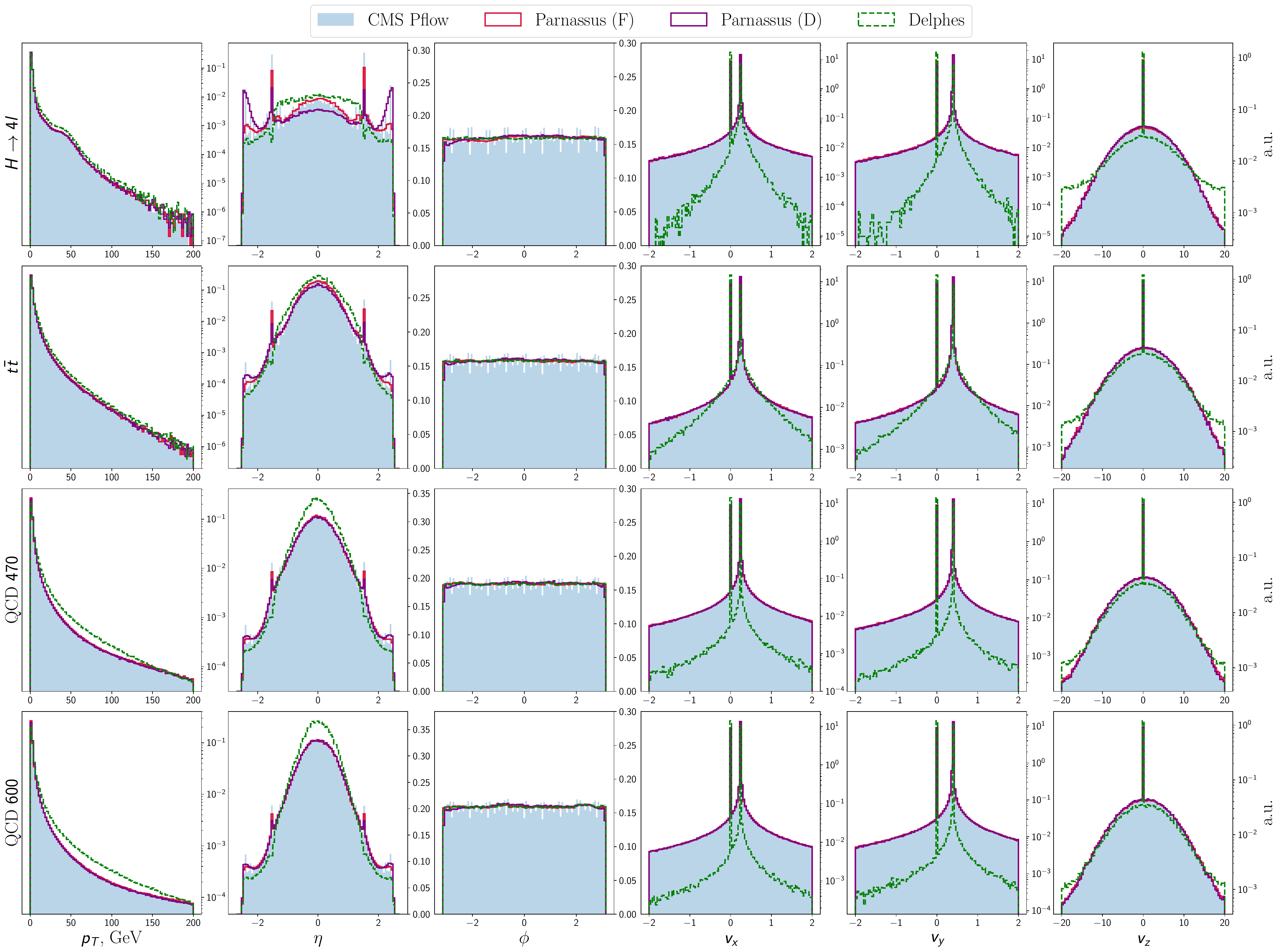}
    \caption{Distributions of particle features for $t\bar{t}$, $H\to 4l$ and QCD datasets.  Rows correspond to different processes while columns correspond to different observables.  All histograms are normalized to unity.}
    \label{fig:all_part}
\end{figure}

The overall spectra of PFCs are presented in~\autoref{fig:all_part}.  The details of the varying detector subsystems are noticeable in the $\eta$ distribution of particles, which is unimodal at particle-level, but develops peaks at transition regions at detector level.  As the detector model in \textsc{Delphes} has the same transition regions as the real CMS detector, the location of these features are well-modeled, even if the fine details of the spectrum at those locations is not as precise as \textsc{Parnassus}. All fast simulation methods average over the small modulations in $\phi$, which are only barely visible even with the fine binning chosen for~\autoref{fig:all_part}.  The diffusion model is more accurate at high $\eta$ compared to the flow model, while the flow model is more accurate at central $\eta$ compared with the diffusion model. The vertex information is used for pileup rejection and flavor tagging -- including this low-level information allows for these quantities to be included in downstream high-level taggers.  \textsc{Delphes} is not designed to precisely model the resolution in $v_x$ and $v_y$, but is tuned to approximately capture the spread in $v_z$.  In contrast, both \textsc{Parnassus} models reproduce all of the properties of the CMS data.  The performance of \textsc{Parnassus} in the first and last rows is strong evidence that the models are learning universal properties of the detector response, as these events are quite different from those seen during training.

\begin{figure}[h!]
    \centering
    \includegraphics[width=0.95\textwidth]{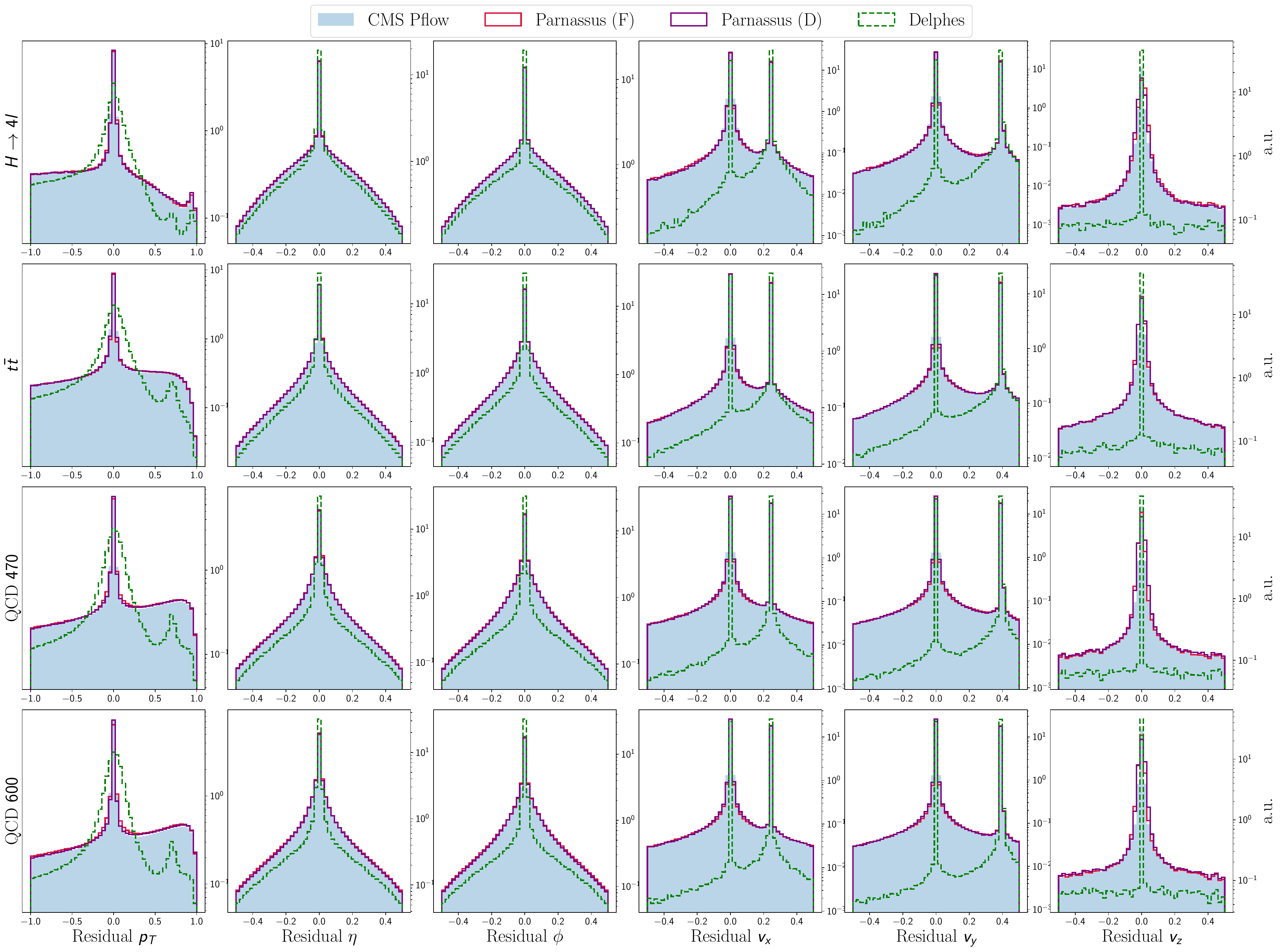}
    \caption{Residual distributions of matched particle features for $t\bar{t}$, $H\to 4l$ and QCD datasets. Rows correspond to different processes while columns correspond to different observables.  All histograms are normalized to unity.}
    \label{fig:all_part_res}
\end{figure}

For PFCs that are matched to GEN particles, the relative residual distributions are presented in~\autoref{fig:all_part_res}.  The particle $p_\mathrm{T}$ residual shows a narrow peak around zero and then a broad shoulder on both sides of this peak.  Since we do not track energy deposits through the detector, it is only possible to match PFC and GEN particles based on geometry and energy.  We interpret the shoulders in the first column of ~\autoref{fig:all_part_res} as actually corresponding to either truly unmatched particles that happen to be ``close'' or cases where a GEN particle created more than one PFC or one PFC is due to more than one GEN particle.  This can result from from material interactions, confusion in reconstruction, or the finite granularity of the detector.  As in the jet case, the resolutions from \textsc{Delphes} tend to be narrower than CMS, while the \textsc{Parnassus} models seem to capture both the bias and variance of the resolution functions, for both in-distribution and out-of-distribution events.  As the tracker and calorimeter granularity are similar in $\eta$ and $\phi$, the resolutions in the two angular coordinates are similar.  The $p_\mathrm{T}$ dependence for the energy resolution is not as prominent as for the angles because of the tradeoff between the tracker resolution that grows with energy and the calorimeter resolution that decreases with energy.

\begin{figure}[h!]
    \centering
    \includegraphics[width=0.95\textwidth]{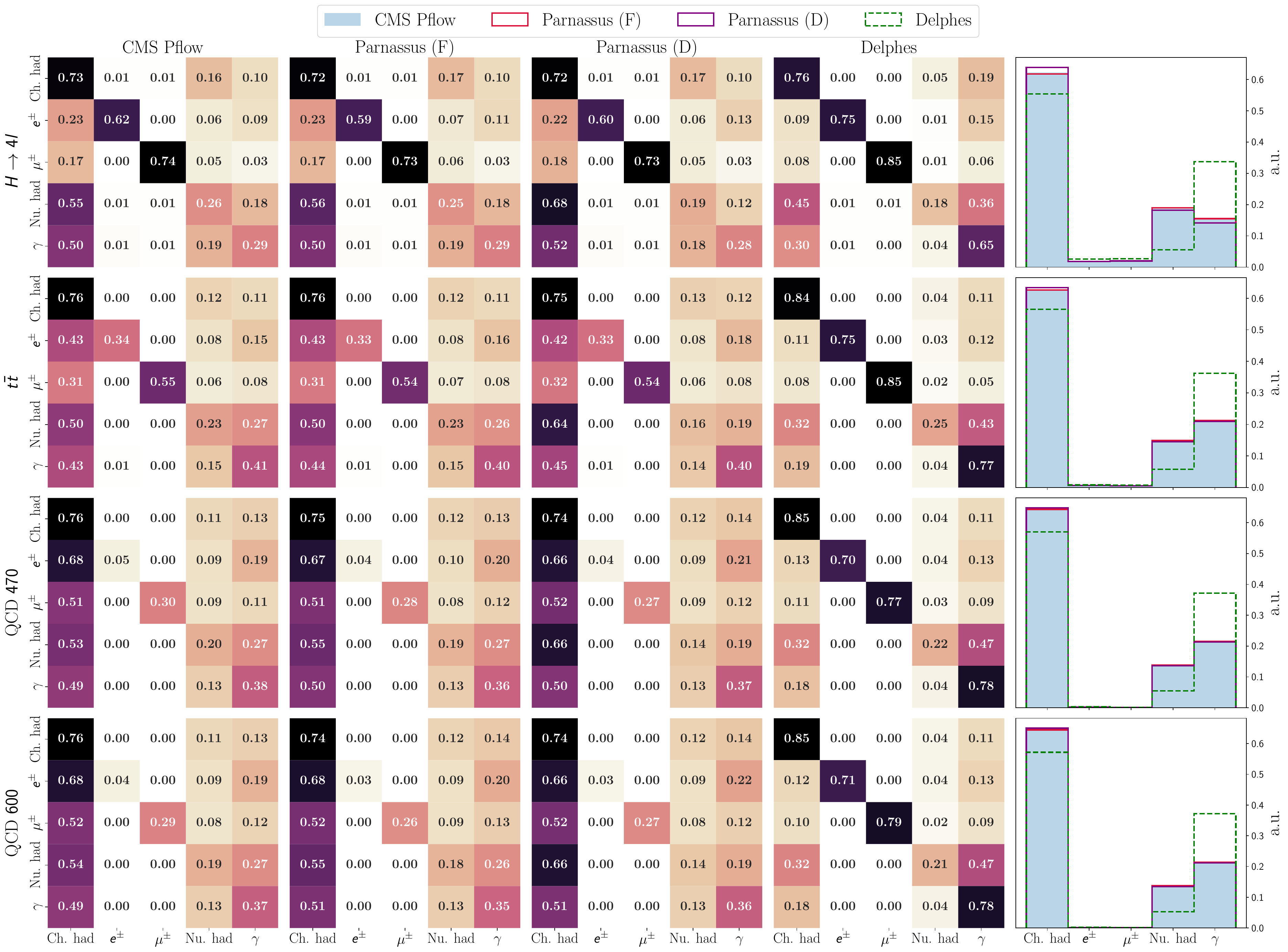}
    \caption{Class confusion matrices of matched particles (first four columns) and all particle distributions (rightmost column) for $t\bar{t}$, $H\to 4l$ and QCD datasets. For the confusion matrices, the truth particle class is on the vertical axis and the reconstructed class is on the horizontal axis.  The frequency distributions are normalized to unity.}
    \label{fig:all_class}
\end{figure}

Next, we show the performance of the particle class reconstruction in ~\autoref{fig:all_class}.  Possible options include charged hadrons (Ch. had), electrons, muons, neutral hadrons (Nu. had) and photons.  The first four columns of the figure depict confusion matrices between truth particles (vertical axis) and reconstructed particles (horizontal axis).  The matching is the same as from~\autoref{fig:all_part} and carries the same limitations (e.g. there is no unique match; charged PFCs can be matched to neutral GENs, etc.).  The goal of the fast simulation methods is to match the CMS performance, not to make the matrix as diagonal as possible.  The \textsc{Parnassus} models reproduce all of the trends well, while the \textsc{Delphes} simulation is much more diagonal than CMS.  The different properties of the four event types are also visible in the matrices.  For example, the probability of an event containing an electron is small except in Higgs and $t\bar{t}$ events, where electrons are part of the hard-scatter.  Interestingly, \textsc{Parnassus} accurately models the muon classification accuracy (about 30\%) even in the QCD events, where the muons are mostly from decays-in-flight from light- or heavy-flavor hadrons.  The last column of ~\autoref{fig:all_class} contains histograms of the frequency for reconstructing particles of a given type.  



These last sets of plots have focused on inclusive PFCs matched to GEN particles.  It is also interesting to see if the performance is the same for particles inside jets (the subject of Ref.~\cite{Dreyer:2024bhs}) and outside jets (`background').  Figures corresponding to the previous figures for these classes of particles are presented in~\autoref{fig:pfcsbrokendown} and~\autoref{fig:pfcsbrokendown_pid}.  While the spectra of in-jet and out-of-jet particles are quite different (higher $p_\mathrm{T}$ and more central inside jets), the \textsc{Parnassus} models are able to match the CMS resolution functions.  The overestimation of the amount of smearing by \textsc{Delphes} is present inside and outside of jets.

Lastly, we explore the properties of PFCs not matched to GEN particles in~\autoref{fig:unmatched_all}.  These PFCs consist of genuine fake particles as well as PFCs not properly associated with their GEN particle from the matching algorithm (e.g. because more than one PFC or more than one GEN belong together).  The fakes are mostly due to pileup (since we lack pileup GEN particles) and not random combinations of hits within the detector.  Compared to matched particles, the unmatched PFCs are softer and more forward.  These trends are captured by both \textsc{Delphes} and \textsc{Parnassus}, with a better level of agreement with CMS from the machine learning models.  For the kinematic properties, this is most prominent at high $p_\mathrm{T}$, where \textsc{Delphes} significantly overestimates the rate in $H\rightarrow 4\ell$ and $t\bar{t}$ events and significantly underestimates the rate in QCD dijet events.  The two \textsc{Parnassus} models are about the same, with slightly better performance from the diffusion model in QCD and at high $|\eta|$ and slightly better performance from the flow model at central pseudorapidity. \textsc{Delphes} is not able to model the vertex distributions for these PFCs, while \textsc{Parnassus} provides an accurate model of their vertex distributions.  By construction, \textsc{Delphes} is not able to produce pure fakes, since it only smears GEN particles and cannot create new particles.  In contrast, the \textsc{Parnassus} models are able to create new, fake particles from scratch.

\section{Conclusion and Outlook}
\label{sec:conclusions}

In this paper, we have extended the \textsc{Parnassus} fast simulation and reconstruction approach to entire events.  This has required us to accommodate much larger particle multiplicities than from single jets.  Furthermore, we now model the particle type and vertex information.  Using the CMS detector as an example, we have shown that \textsc{Parnassus} is an accurate surrogate model for simulation and reconstruction, with excellent performance even on physics processes outside of the training dataset.  The training of \textsc{Parnassus} is fully automated and the inference is purely Pythonic and GPU-compatible.   

Our vision is that \textsc{Parnassus} will provide an accurate, portable and low-barrier-to-entry framework for fast simulation and reconstruction across particle physics experiments. This paper marks the first setup that is ready for deployment.  As next steps, we will pursue three directions. First, we will continue to improve the accuracy, precision, stability, and speed of the machine learning models.   The current level of accuracy is sufficient to go beyond the capabilities of \textsc{Delphes} and it may one day be able to provide a direct complement to collaboration-internal fast simulation and reconstruction tools.  Second, we will build a software interface to make it easy for users to connect particle-level Monte Carlo event generators with their own detector-level analysis, including calibrations.  Lastly, we will facilitate the training of a suite of detector models (starting with the two CMS ones in this paper), so that users can study diverse particle physics experiments at colliders and beyond.

\section*{Data and Code Availability}

The code for this paper can be found at \url{https://github.com/parnassus-hep/cms-flow-evt}. The postprocessed CMS dataset and the generated \textsc{Delphes} dataset can be found on Zenodo at \url{https://zenodo.org/records/15083495}.

\section*{Acknowledgments}
VM and BN are supported by the U.S. Department of Energy (DOE), Office of Science under contract DE-AC02-05CH11231.
EG, ED, and DK, are supported by the
Minerva Grant 715027, and the Weizmann Institute for Artificial Intelligence grant
program Ref 151676.

\clearpage

\appendix
\section{Additional Plots}

\begin{figure}[h!]
    \centering
    \begin{subcaptionblock}{.45\textwidth}
        \includegraphics[width=0.95\linewidth]{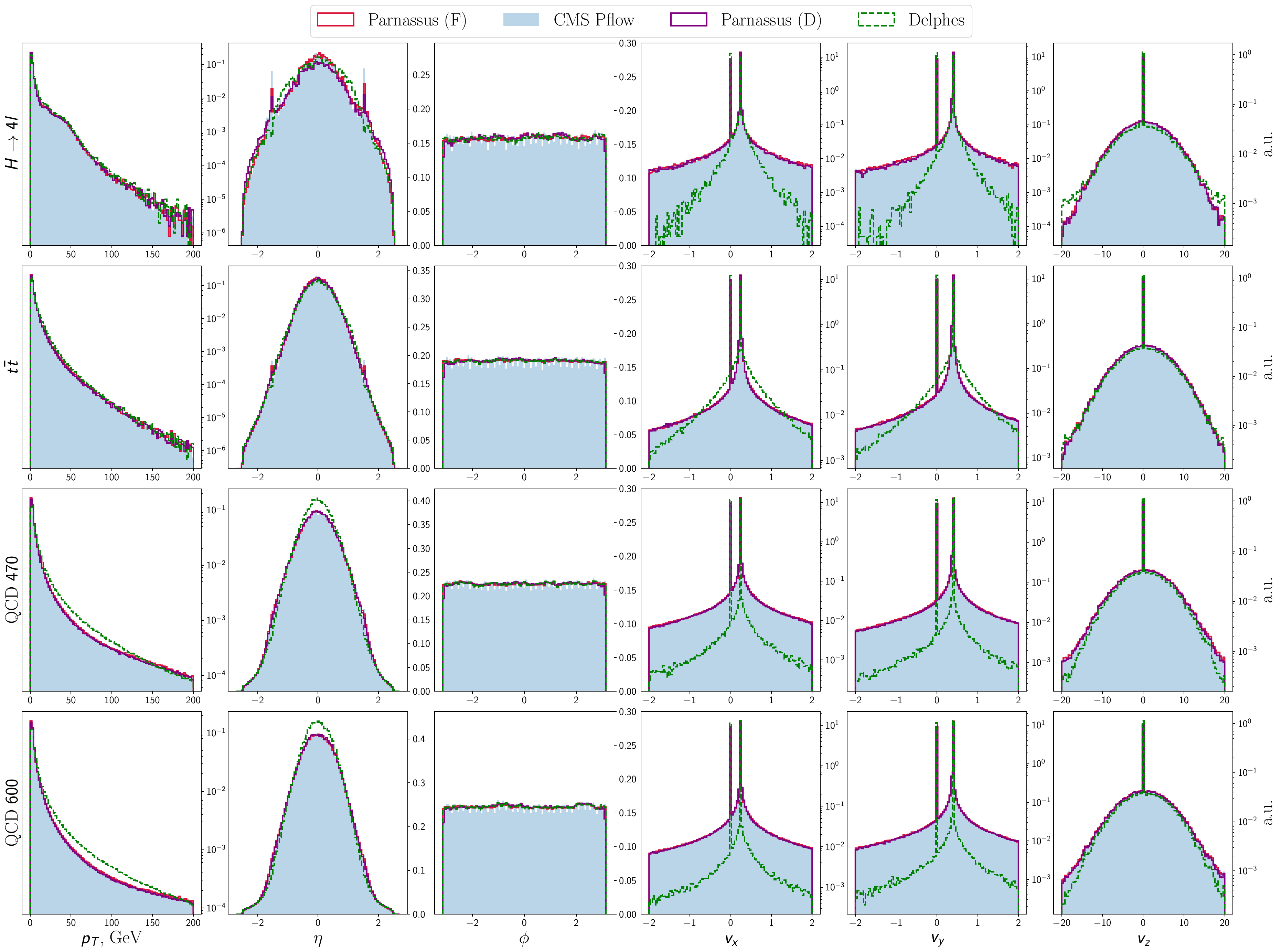}
        \caption{\footnotesize Distributions of matched to jets particle features}
    \end{subcaptionblock}
    \begin{subcaptionblock}{.45\textwidth}
        \includegraphics[width=0.95\linewidth]{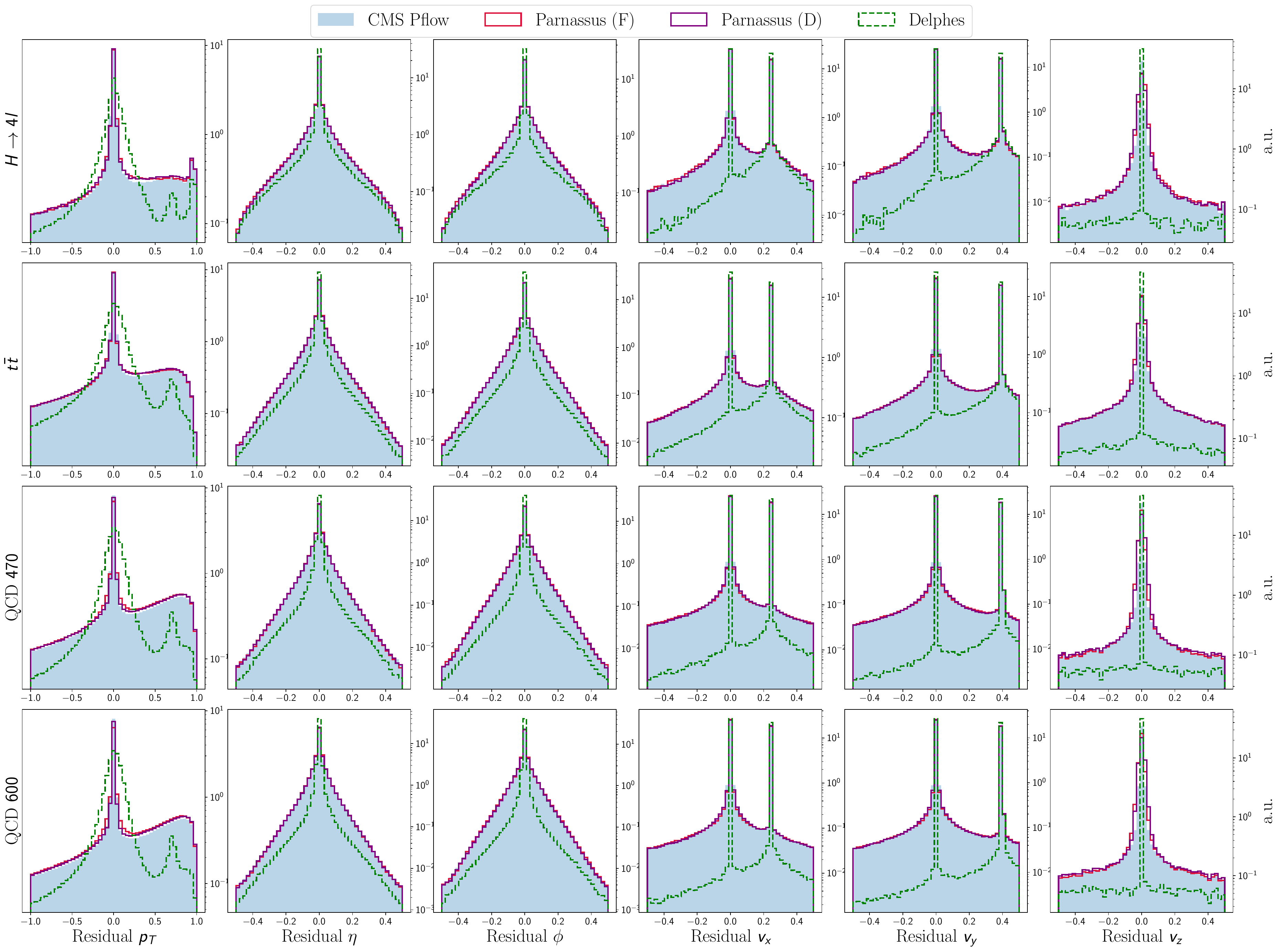}
        \caption{\footnotesize Residual distributions of matched to jets particle features}
    \end{subcaptionblock}
    \begin{subcaptionblock}{.45\textwidth}
        \includegraphics[width=0.95\linewidth]{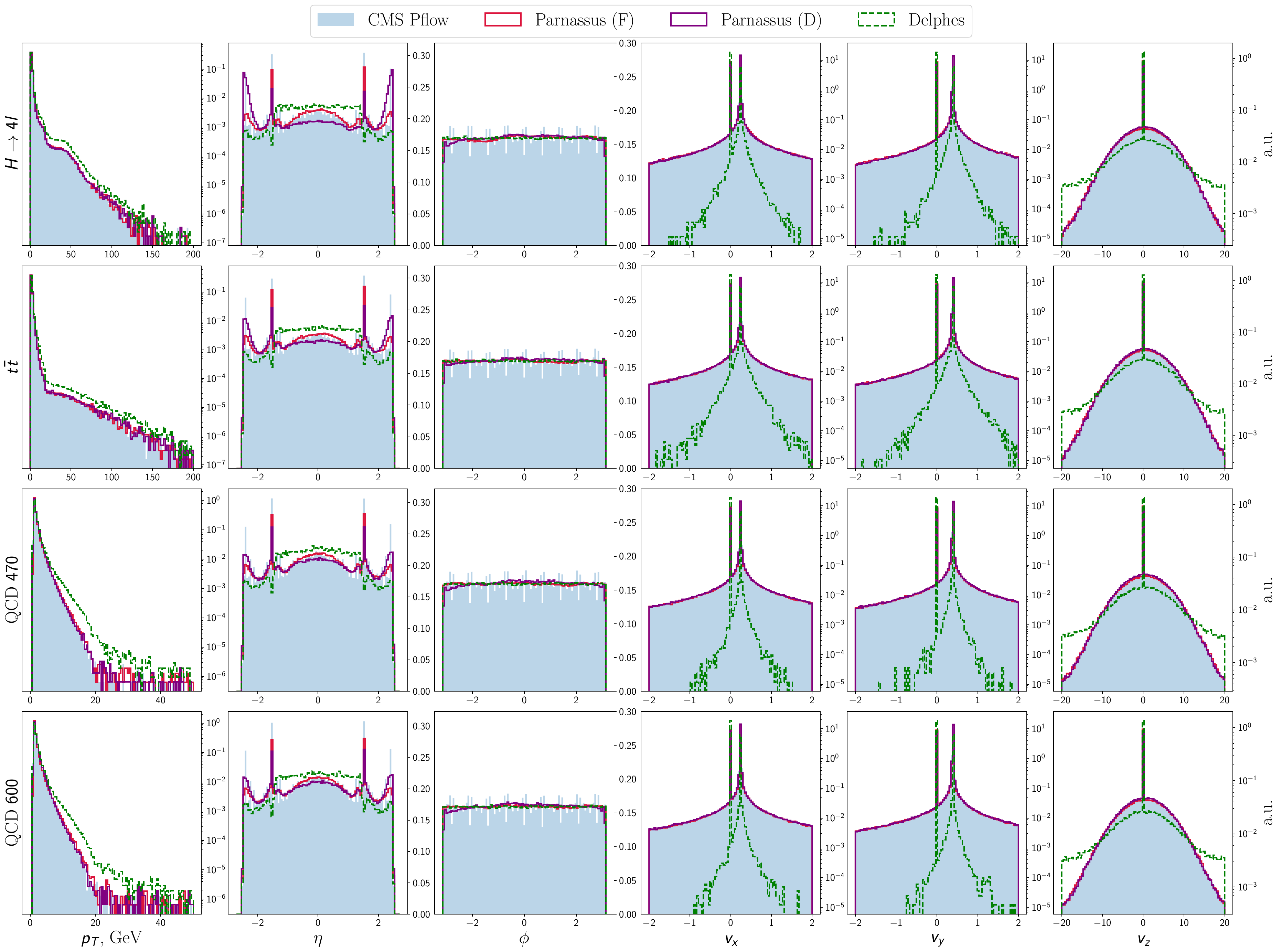}
        \caption{\footnotesize Distributions of background particle features}
    \end{subcaptionblock}
    \begin{subcaptionblock}{.45\textwidth}
        \includegraphics[width=0.95\linewidth]{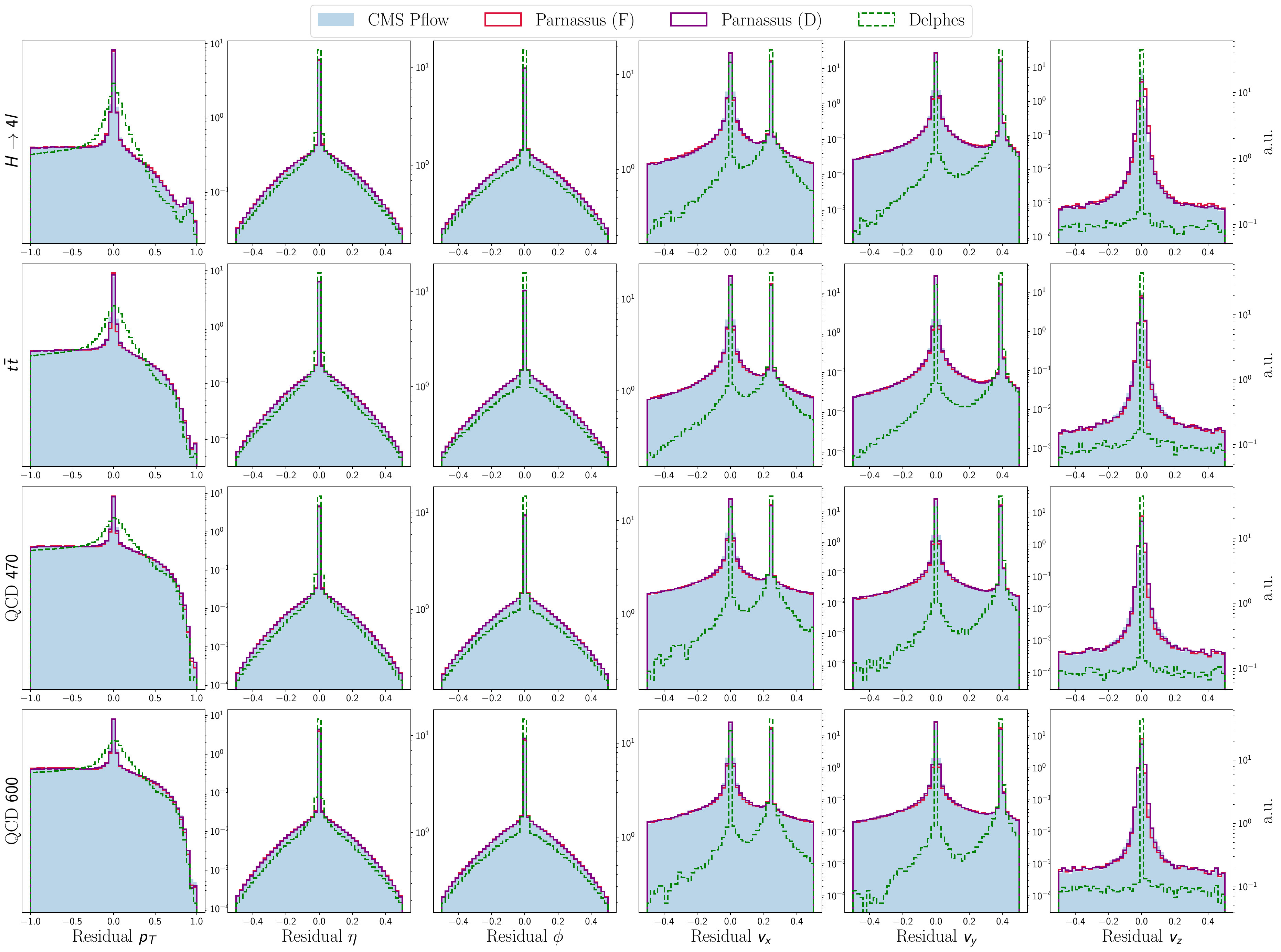}
        \caption{\footnotesize Residual distributions of background particle features}
    \end{subcaptionblock}
    \caption{Kinematic properties of particles in jets (top) and out of jets (bottom).}
\label{fig:pfcsbrokendown}
\end{figure}

\begin{figure}[h!]
    \centering
    \begin{subcaptionblock}{.45\textwidth}
        \includegraphics[width=0.95\linewidth]{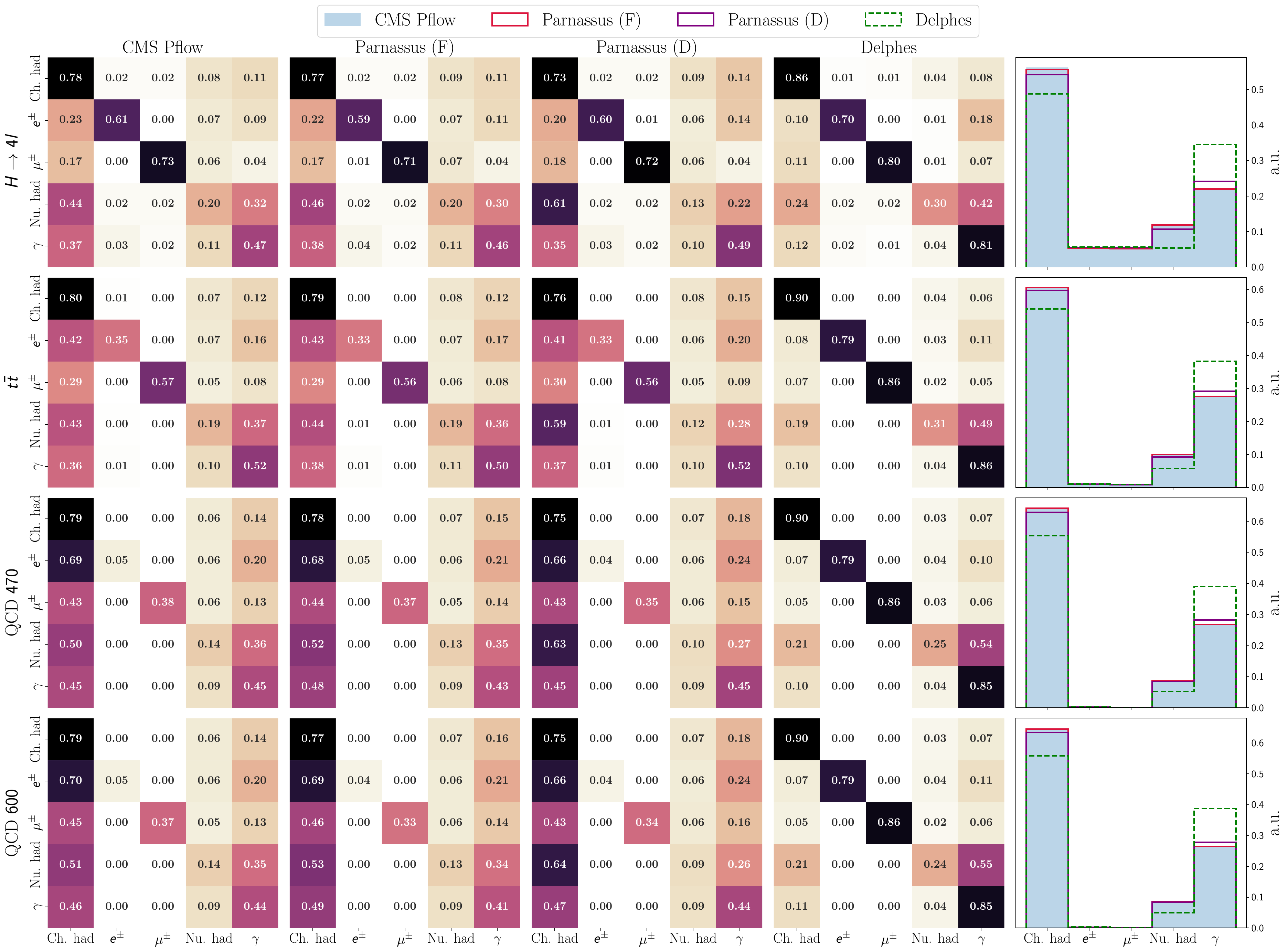}
        \caption{\footnotesize Class confusion matrices of matched to jets particles.}
    \end{subcaptionblock}
    \begin{subcaptionblock}{.45\textwidth}
        \includegraphics[width=0.95\linewidth]{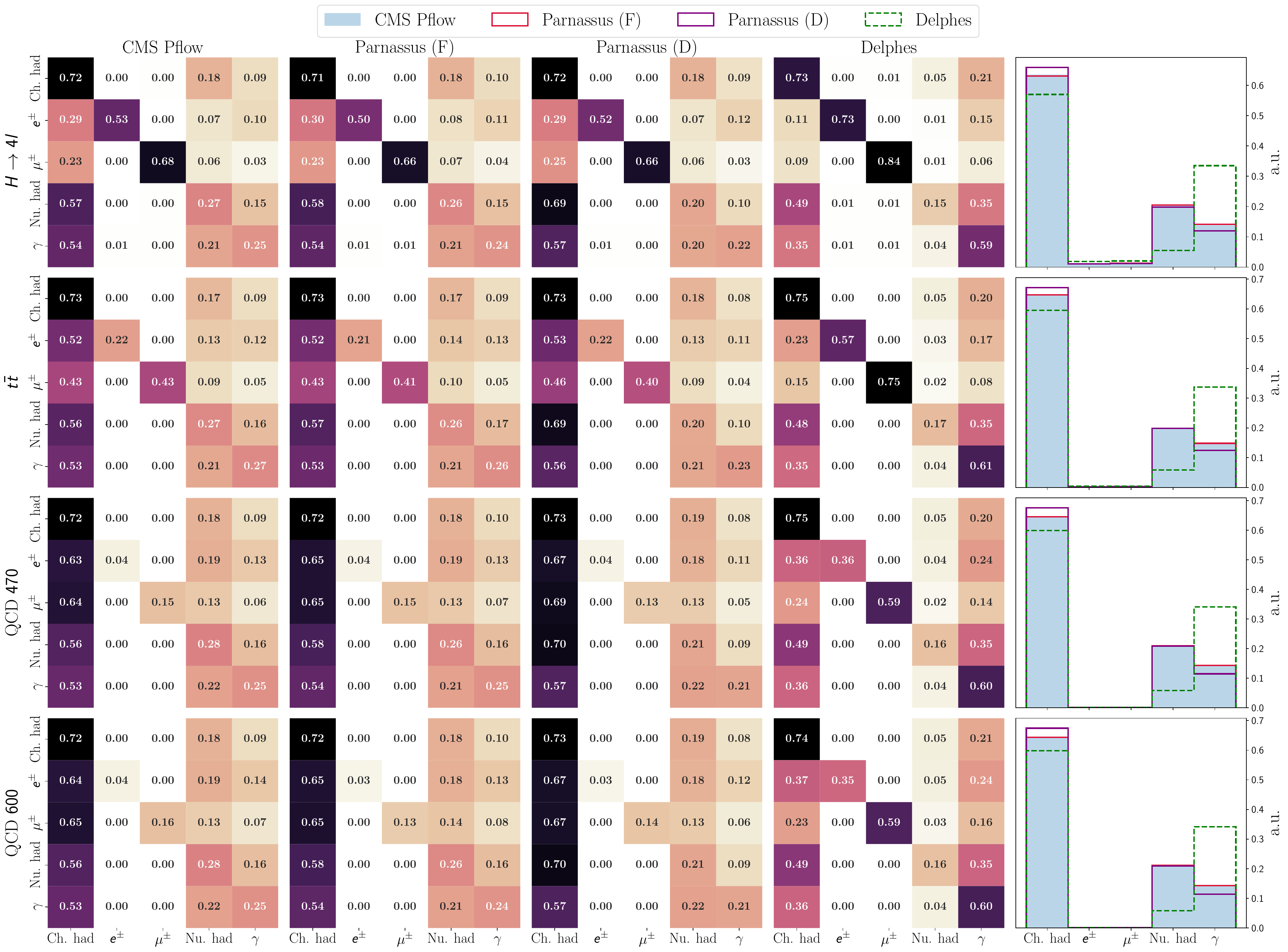}
        \caption{\footnotesize Class confusion matrices of background particles.}
    \end{subcaptionblock}
        \caption{Particle identification performance for in-jet (left) and out-of-jet (right) particles.}
    \label{fig:pfcsbrokendown_pid}
\end{figure}

\begin{figure}[h!]
    \centering
    \begin{subcaptionblock}{.45\textwidth}
        \includegraphics[width=0.95\textwidth]{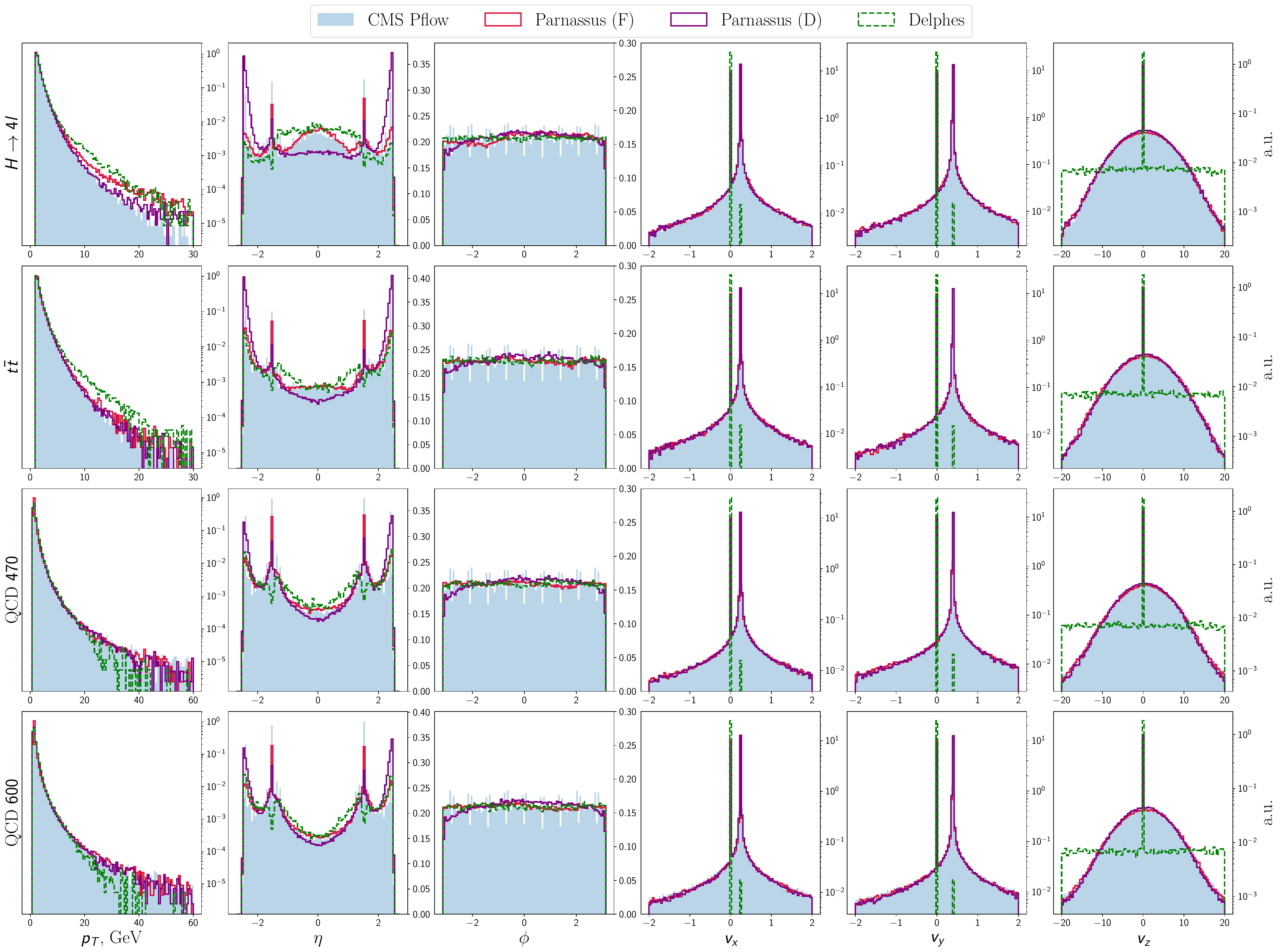}
        \caption{\footnotesize All particles.}
    \end{subcaptionblock}
    \begin{subcaptionblock}{.45\textwidth}
        \includegraphics[width=0.95\textwidth]{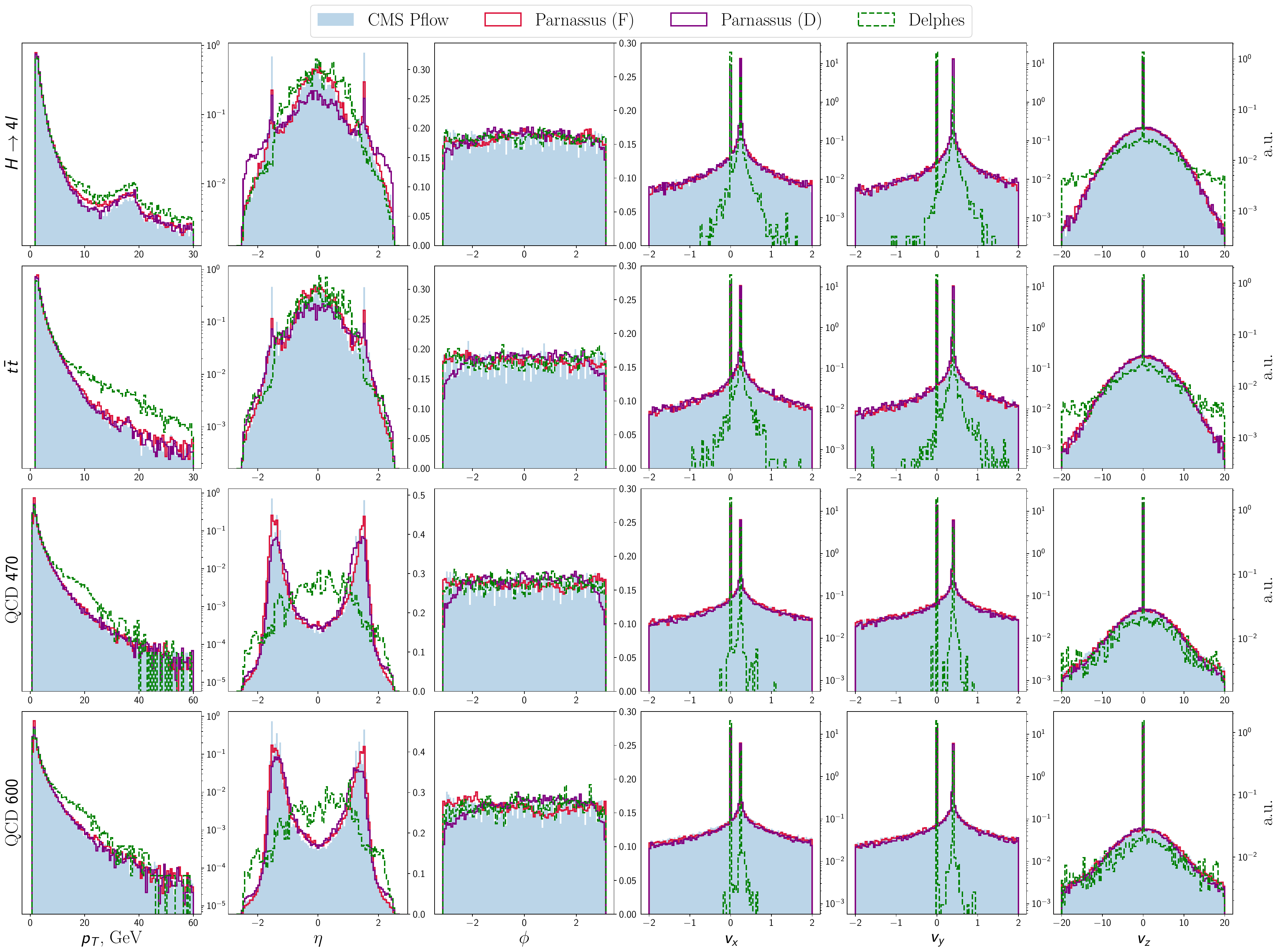}
        \caption{\footnotesize Particles matched to jets.}
    \end{subcaptionblock}
        \caption{Distributions of PFCs not matched to GEN particles.}
    \label{fig:unmatched_all}
\end{figure}


%
\clearpage

\bibliography{HEPML,other,fastml}
\bibliographystyle{apsrev4-1}

\clearpage
\onecolumngrid

\appendix

\end{document}